\documentstyle[12pt,epsf]{article}
\newtheorem{thm}{Theorem}

\newtheorem{lem}{Lemma}
\newtheorem{df}{Definition}
\newtheorem{cor}{Corrolary}

 \newcommand{\Integer}{\:\mbox{\rm \sf Z} \hspace{-0.82em} \mbox{\rm \sf Z}\,}
  
 \newcommand{\Real}{\mbox{\rm I \hspace{-0.82em} R}}
 
 \newcommand{\Complex}{
        \mbox{\rm C \hspace{-1.16em} \raisebox{-0.018em}{\sf l}}\;}
 \newcommand{\Z}{\Integer}

 \newcommand{\bd}{\begin{df}}
 \newcommand{\bt}{\begin{thm}}
 \newcommand{\bl}{\begin{lem}}
 \newcommand{\ed}{\end{df}}
 \newcommand{\be}{\begin{equation}}
 \newcommand{\ee}{\end{equation}}
 \newcommand{\bi}{\begin{itemize}}
 \newcommand{\ei}{\end{itemize}}
 \newcommand{\et}{\end{thm}}
 \newcommand{\el}{\end{lem}}
 \newcommand{\bs}{\bigskip}
 \newcommand{\ms}{\medskip}
 \newcommand{\eb}{\hfill$\Box$}
 \newcommand{\tl}{{\cal T}}
 \newcommand{\tr}{\mbox{\rm tr}}
 
 \newcommand{\Ch}{\chi^{}}
 \newcommand{\A}{{\cal A}}
 \newcommand{\tilx}{pointed tiling}
 \newcommand{\tilxx}{doubly pointed tiling}
 \newcommand{\mi}{pattern}
 \newcommand{\mix}{pointed pattern class}
 \newcommand{\mixx}{doubly pointed pattern class}
 \newcommand{\miii}{pattern class}
 \newcommand{\ti}{tile}

 \newcommand{\B}{{\cal B}}

 \newcommand{\Om}{\Omega}
 \newcommand{\om}{\omega}
 
 \newcommand{\pf}{{\cal P}_{\Sigma}}
 \newcommand{\AF}{{\cal A}_{\Sigma}}
 \newcommand{\pfw}{Perron-Frobenius-eigen\-value}
 \newcommand{\pfv}{Perron-Frobenius-ei\-gen\-vec\-tor}
 \newcommand{\Gr}{{\cal R}}
 \newcommand{\Gru}{\Gamma}
 \newcommand{\So}{Schr\"odinger operator}
 \newcommand{\cm}{connectivity matrix}
 \newcommand{\sst}{substitution}

 \newcommand{\fl}{substitute}
 \newcommand{\nc}{non commutative}
 
 \newcommand{\saum}{border}
 \newcommand{\CA}{$C^*$-algebra}
 \newcommand{\K}{{\cal K}}

\newcommand{\V}{{\cal V}}

\newcommand{\si}{stably isomorphic}
\newcommand{\mk}{decorat}
\newcommand{\pe}[1]{$#1$-facet}
\newcommand{\faden}{line}

 \newcommand{\Gs}{{\cal S}}
\newcommand{\kt}[1]{X(#1)}
\newcommand{\kT}{\kt{\tc}}

\newcommand{\mT}{{\cal M}}
\newcommand{\mtxx}[1]{{\cal M}_{\rm I\!I}(#1)}
\newcommand{\mtx}[1]{{\mTxx(#1)}^0}
\newcommand{\mTxx}{{\cal M}_{\rm I\!I}}
\newcommand{\mTx}{{{\mTxx}^0}}

\newcommand{\mTzw}{{\mTxx}_{2,\neq}}

\newcommand{\mTei}{{\mTxx}_{1}}
\newcommand{\tc}{T}

\newcommand{\C}{{\cal C}}
\newcommand{\U}{{\cal U}}
\newcommand{\ein}{\preceq}
\newcommand{\einp}{\dot{\preceq}}
\newcommand{\cmp}{\cp}
\newcommand{\ac}{\rhd}

\newcommand{\mul}{}
\newcommand{\cp}{\vdash}
\newcommand{\act}{\cdot}
\newcommand{\AG}{almost-groupoid}
\newcommand{\rad}{\mbox{\rm rad}}
\newcommand{\erz}[1]{\langle{#1}\rangle}
\newcommand{\erzC}{\erz{\C}}

\newcommand{\im}{\mbox{\rm im}\,}
\newcommand{\isg}{{\cal ISG}}
\newcommand{\asg}{{\cal ASG}}
\newcommand{\N}{{\cal N}}
\newcommand{\svarphi}{\hat{\varphi}}
\newcommand{\sva}{\svarphi}
\newcommand{\spsi}{\hat{\psi}}

\newcommand{\srho}{\hat{\rho}}
\newcommand{\KK}{{\cal L}}
\newcommand{\F}{{\cal F}}
\newcommand{\mb}{\upsilon}
\newcommand{\Laml}{\Lambda_\mb}
\newcommand{\kpl}{{\kappa_\mb}}
\newcommand{\Ql}{Q_\mb}
\newcommand{\PfL}{{\cal P}_{\Laml}}
\newcommand{\Kl}{K^{(l,\mb)}}
\newcommand{\KKl}{\KK^{(l,\mb)}}
\newcommand{\laml}{\lambda_\mb}
\newcommand{\eg}{\alpha}
\newcommand{\Rh}{{Rh}}
\newcommand{\crho}{\check{\rho}}
\newcommand{\fp}[1]{\grave{#1}}
\newcommand{\lp}[1]{\acute{#1}}
\newcommand{\bp}[1]{\check{#1}}
\newcommand{\trho}{\rho_t}

 \title{
 \bf The Local Structure of Tilings and their Integer Group of Coinvariants
   \vspace{1.5em}}
 \author{Johannes Kellendonk}
 \date{Department of Mathematics, King's College London,\\
       Strand, London WC2R 2LS\\
        \vspace{.2em}
       {\small E-mail: johannes@mth.kcl.ac.uk}}

\begin{document}
 \maketitle

 \begin{abstract}
\noindent
The local structure of a tiling is described in terms of a
multiplicative structure on its pattern classes.
The groupoid associated to the tiling is derived from this structure
and its integer group of coinvariants is defined.
This group furnishes part of the $K_0$-group of the groupoid
$C^*$-algebra for tilings which reduce to decorations of
$\Z^d$. The group itself as well as the image of its state is
computed for \sst\ tilings in case the \sst\ is locally invertible and
primitive. This yields in particular the set of possible gap labels
predicted by $K$-theory for Schr\"odinger operators describing the particle
motion in such a tiling.

 \end{abstract}
 \begin{flushright}
 \parbox{12em}
  { \begin{center}
      KCL-TH-95-6
 \end{center} }
 \end{flushright}

\newpage

\bibliographystyle{unsrt}

\section*{Introduction}

\addcontentsline{toc}{section}{\bf Introduction}

Finding topological objects to characterize or even
classify tilings could be seen as a
major motivation to study tilings. This article is less
ambitious but already has the taste of it in that it is largely
devoted to the study of such an object: the integer group of coinvariants
associated to the tiling and the range of a state on it. But it takes
its stimulation from two areas of application. One is the
$K$-theoretical gap labelling of \So s describing the motion of a particle
in a tiling (and potentially many more features of solids for which a model
by a tiling is appropriate), and the other concerns topological dynamical
systems on the Cantor set. Let us briefly put these two areas into
context.

1) Tilings furnish (discrete) models of solids which may not be
periodic as for instance
quasicrystals.
A typical question arrising is that after the nature of the
spectrum of a particle moving in the tiling (e.g.\ a phonon or an electron).
The corresponding \So\ lies in the algebra of observables but its
diagonalization is for non periodic tilings to difficult to carry out
at present. To obtain at least some
qualitative description of the spectrum Bellissard proposed the $K$-theoretical
gap labelling \cite{Be4}. This gap labelling
requires the computation of a topological
invariant of a \CA\ which may be taken to be the algebra of observables,
namely of the range of a tracial state on its $K_0$-group.
The algebra of observables coincides with the algebra associated to the
tiling and the tracial state comes from a trace on it. The $K_0$-group
contains the integer
group of coinvariants (at least for a large class of tilings)
and the range of the tracial state equals its range on the coinvariants.
This range is a countable subgroup of $\Real$ which contains the possible
labels
of gaps in the spectrum of a \So, i.e.\ its elements are the gap labels
predicted by $K$-theory.

As it is determined by
a topological invariant of the \CA\ associated to the tiling
the $K$-theoretical gap labelling will be of use if the
nature of the tiling is incorporated in the operator in a strong enough way
so that
all values of the above group (lying between $0$ and $1$) correspond to gaps.
Although the quantification of the latter property is in general (in particular
in higher dimensions)
an unsolved problem we find it still worth persuing as there are no
analytical alternatives in higher dimensions so far.

2) A certain class of tilings, the decorations of $\Z^d$, yield
$d$-dimensional topological dynamical systems of the Cantor set.
Such a topological dynamical system is given by $d$ commuting homeomorphisms,
and a tiling which consists of decorated
$d$-dimensional unit cubes touching nicely at their faces yields an example,
the $d$ independent shifts yielding commuting homeomorphisms of the hull of the
tiling. In one dimension the study of the $K$-group of the associated
\CA\ (the one usually associated to the dynamical system coincides with the
one associated to the tiling)
turned out to be very fruitful. The ordered $K_0$-group with order unit,
which coincides with the integer group of coinvariants as an unordered group,
furnishes an invariant classifying the classes
of strong orbit equivalent dynamical systems \cite{HPS,GPS}.
In higher dimensions the interpretation of the $K$-group in terms of
the properties of dynamical systems is yet less clear but it was recently
realized that the (unordered) $K$-groups may be decomposed into cohomology
groups of the group $\Z^d$ \cite{FoHu}. In particular the group of
coinvariants furnishes part of the $K_0$-group thus motivating its computation
which is carried out below for tilings which allow for an invertible \sst.
In turn, most of the analysis done for decorations of $\Z^d$ remains valid
for a much larger class of tilings including the Penrose tilings, as will
be shown below using the concept of reduction.\bs

Although the above applications are formulated in the framework of
$K$-theory of \CA s we shall not emphasize the \CA ic aspects in this
article since the topological invariant discussed here may be defined purely on
the level of spaces, i.e.\ here of groupoids.
The article is organized as follows.

In the first section the groupoid associated to a tiling is described.
We proceed along lines sligthly different from \cite{Ke2}
by first introducing a multiplicative structure on the pattern classes
of the tiling. This multiplicative structure is almost a groupoid, and
using the notion of the radius of a pattern class one obtains
the groupoid associated to the tiling in a way which makes its
local nature transparent. In contrast to \cite{Ke2} this groupoid is only
principal if the tiling is not periodic whereby we gain that it yields
the right groupoid \CA\ in the periodic case, too.
We then define the integer group of coinvariants associated to the tiling.
Naturally, local manipulations, i.e.\ deformations of patterns, appear
as maps which
are close to being homomorphims of the introduced \AG s.
In fact, a \sst\ can be algebraically defined as a particular kind of such a
homomorphism. We discuss the concept of reduction and the notion of a
decoration of $\Z^d$ and show that many tilings occuring in the description
of quasicrystals reduce to decorations of $\Z^d$.

Section 2 is devoted to the \CA ic characterization and contains some
details about the $K$-theoretical gap labelling. We first recall the
definition of the algebra associated to the tiling. The main result
needed concerns tilings which reduce to decorations of $\Z^d$ and states
that their \CA\ is stably isomorphic to a crossed product with $\Z^d$.
In particular the
$K$-theory of such a tiling is that of a topological dynamical system.

The third section, which is kept independent of
sections 1.4 and 2, is restricted
to tilings which allow for an invertible \sst. Geometrically such a \sst\
may be obtained from an deflation but the analysis remains purely algebaic.
The existence of an invertible \sst\ reflects in the
possibility of obtaining an injective coding of the tiling in terms of the
paths over a graph. This is technically the
basis for the computation of the integer group of coinvariants and the range
of its state.
For one dimensional \sst s similar approaches have been successfully
carried out, both,
for the range of the state \cite{Q,BBG} as well as for the group itself
\cite{For,Hos}. In higher dimensions results regarding the
range of the state have been obtained under the restriction that
the \sst\ forces its border \cite{Ke2}.
The results obtained here are independent of the dimension of the tiling,
the only requirements being that the \sst\ is (locally) invertible
and, for the determination of the range of the state, primitive.
In fact, we obtain a family of codings and graphs for a tiling which
for \sst s which force their border all coincide and have
the \sst\ matrix as
connectivity matrix but are more complicated in the general case.
As applications we discuss the Thue-Morse \sst\ sequence and the
Penrose tilings.


\section{Tilings and groupoids}

Groupoids and \AG s play a central role in the description of tilings.
They carry a structure which generalizes a group structure in so far as
the multiplication of two elements $x$ and $y$ is only defined provided
$(x,y)$ lie in a specified set. We shall indicate this by $x \cp y$.
As a general reference to groupoids which includes groupoid-\CA s we
use \cite{Ren}.

\subsection{Almost-groupoids}

\bd
An \AG\ is a set $\Gru$ together with a subset
$\Gru^{\cmp}\subset\Gru\times\Gru$ of composable elements (we write
$x\cp y$ for $(x,y)\in\Gru^{\cmp}$)
and a product map $x\cp y\mapsto xy$:
$\Gru^{\cmp}\rightarrow\Gru$ and an inversion 
map $x\mapsto x^{-1}$: $\Gru\rightarrow\Gru$, such that the
following relations hold:
\bi
\item[I1] $(x^{-1})^{-1} = x$,
\item[I2] $x\cp y$ implies $y^{-1}\cp x^{-1}$, in which case
 $(xy)^{-1}=y^{-1}x^{-1}$,
\item[I3] $x\cp x^{-1}$ and $x\cp x^{-1}x$ and $xx^{-1}x=x$,
\item[A]
 $(x\cp y$ and $xy\cp z)$ whenever $(x\cp yz$ and $y\cp z)$,
 in which case $(xy)z = x(yz)$.
\ei
\ed
The elements of $\Gru^0=\{x^{-1}x|x\in\Gru\}$ are called units.
We will make frequent use of the maps $L,R:\Gru\rightarrow\Gru^0$
(often denoted $r,d$)
defined by
\be
L(x) = xx^{-1} \quad\quad R(x)=x^{-1} x.
\ee
An \AG\  is called principal
if the map $(L,R):\Gru\to\Gru^0\times\Gru^0$ is injective.
It defines orbits in the space of units: $u$ and $v$ are called
orbit equivalent, $u\sim_{o} v$, whenever there is an $x$ such that
$L(x)=u$ and $R(x)=v$.
Note that $x\cp y$ is equivalent to $x\cp yy^{-1}y$ and $yy^{-1}\cp y$
so that it is by $A$ equivalent to $x\cp yy^{-1}$. In particular,
$R(x)=L(y)$ implies 
$x\cp y$.
\bd
A groupoid is an \AG\ for which cancelation holds, i.e.
\bi
\item[C]
$x\cp y$ and $x\cp z$ imply $y= z$.
\ei
\ed
This definition of a groupoid is equivalent
to the usual one as e.g.\ presented in \cite{Ren}.
It is not difficult to show that the cancelation axiom C indeed
implies that
\bi
\item
$x\cp y$ implies
$R(x)=L(y)$,
\item
$x\cp y$ and $y\cp z$ imply  $x\cp z$.
\ei
In particular the above implications do not have to hold for \AG s
and associativity is only guaranteed provided all
multiplications are defined.
The lack of cancelation may be put into an order.
\bd
The order of an \AG\ is defined by
\be
x\ein y\quad\mbox{iff}\quad x\cp y^{-1}\quad\mbox{and}\quad xy^{-1}=yy^{-1}.
\ee
\ed
A topological \AG\ is an \AG\ which carries a topology
such that the product and the inversion map are continuous,
$\Gru^\cp$ carrying the relative topology.
A (locally compact) groupoid is called $r$-discrete if $\Gru^0$ is open.

A homomorphism of \AG s is a map which maps composable elements resp.\ units
onto composable elements resp.\ units
and preserves multiplication and inversion.
In particular it preserves the order.
A homomorphism of topological \AG s is a continuous homomorphism
of \AG s.
\bd
An inverse semi-group is an \AG\ for which $\Gru^{\cmp}=\Gru\times\Gru$.
\ed
Any \AG\ can be made into an inverse semi-group by introducing an extra
element $0$ and extending the multiplication to any
$(x,y)\in\Gru\times\Gru$ by
\be
x y = \left\{ \begin{array}{ll}
xy & \mbox{if $x\cp y$} \\
0 & \mbox{else}
\end{array} \right.
\ee
and $0 x = x  0 = 0$, $0 0 = 0$. $0$ is its own inverse and
greater than any other element.
Conversely we could see an \AG\ as an inverse semi-group with an element
$0$ satisfying $0 x = x  0 = 0$ for all $x$.
However, homomorphisms of \AG s may in general not be extended to
homomorphisms of inverse semi-groups and vice versa.
The difference which makes homomorphisms of inverse semi groups unsuitable
for the purposes of this article
is that for a homomorphism $\varphi$ of an \AG,
even if $\varphi(x)\cp \varphi(y)$, it has only to satisfy
$\varphi(x)\varphi(y)=\varphi(xy)$ if $x\cp y$.\bs

{\em Examples:}
1) The principal groupoid which will be associated to a tiling is
given by an equivalence relation, i.e.\
its elements are pairs of equivalent elements of some set $X$.
Multiplication is only defined for pairs $(x,y),(x',y')$ if $x'=y$ and then
given by $(x,y)(y,z)=(x,z)$, and inversion by
$(x,y)=(y,x)^{-1}$.
The topology of the groupoids in question need in
general not to coincide with the relative topology from $X\times X$.

2) Sometimes the equivalence may be expressed as orbit equivalence under
the (right) action of a group $S$ acting on $X$: $x\sim y$ whenever
$\exists s\in S:y=x\cdot s$.
This leads to the consideration of another kind of groupoid which is called
transformation group \cite{Ren}.
Its space is the Cartesian product of $X$ with $S$ (here always considered
to carry the product topology) and the groupoid
structure is defined by $(x,s)(x',t)=(x,st)$ provided $x'=x\cdot s$ and
$(x,s)^{-1}=(x\cdot s,s^{-1})$.
We write it as $X\times_\alpha S$, $\alpha$ indicating the action. It
may be viewed as the groupoid defined by orbit equivalence
only if $S$ acts freely on $X$.

3) Any groupoid $\Gru$ defines an inverse semi-group $\isg(\Gru)$
consisting of so-called $\Gru$-sets. A $\Gru$-set is a subset of
$\Gamma$ for which the restrictions of $L$ and $R$ to it are both injective.
If $X,Y$ are such $\Gru$-sets (which may be empty)
then $XY:=\{xy|x\in X,y\in Y,x\cp y\}$
and $X^{-1}:=\{x^{-1}|x\in X\}$. This multiplication of sets is not only
for $\Gru$-sets defined. We will make frequent use of it.

\subsection{Tilings}

Borrowing the terminology for $d=2$ from \cite{GrSh}
a $d$ dimensional tiling is a (countable) family of
 closed subsets of $\Real^d$, called the
\ti s, which cover $\Real^d$, overlap at most at their boundaries and may
carry an additional decoration, e.g.\ to break symmetries.
We restrict the possible shape of \ti s through the requirements that
they are of finite size, i.e\ they fit inside an $r$-ball for finite $r$,
and they are the closure of their interior.

We shall consider the geometrical objects as
equivalence classes under translations in $\Real^d$ (but not under rotations
or reflections).
A tiling or its class $T$ defines an \AG\ as follows.

Consider the set 
of \miii es (with resp.\ to translation) of $T$
where a pattern of a tiling is given by a finite subset of its \ti s.
A \miii\ $M$ together with two tiles chosen from it (pointed out)
will be called a \mixx. It will be denoted by $M_{xy}$ where $x$ is the first
and $y$ the second pointed tile. The set of all \mixx es of $T$ will be
denoted by $\mtxx{T}$ or simply by $\mTxx$ if the tiling is clear from the
context.

$\mTxx$ carries an order structure: $M_{x_1x_2}\einp N_{y_1y_2}$ iff
the \miii\ of $N_{y_1y_2}$ contains the \miii\ of $M_{x_1x_2}$ at such a
position that \ti s $x_1$ resp.\ $x_2$ become $y_1$ resp.\ $y_2$, c.f.\ figure.

\epsffile[0 0 430 110]{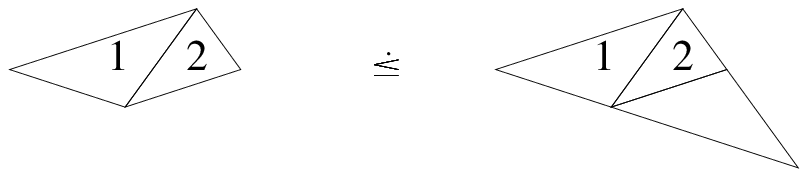}

Moreover $\mTxx$ carries
the structure of an \AG\ whose order structure coincides with the one just
introduced:
Two elements $M_{x_1x_2}$, $N_{y_1y_2}$ are composable iff there is
an $L_{z_1z_2}$ and a third tile $z$ of $L$ such that
$M_{x_1x_2}\einp L_{z_1z}$
and $N_{y_1y_2}\einp L_{zz_2}$.
We then define the product
\be
M_{x_1x_2}\mul N_{y_1y_2} = \min\{L_{z_1z_2}|
\exists z\in\kt{L}
:M_{x_1x_2}\einp L_{z_1z}, N_{y_1y_2}\einp L_{zz_2}\}
\ee
where the minimum is taken with respect to $\einp$ and $\kt{L}$ denotes
the tiles of the pattern class $L$. E.g.:

\epsffile[0 0 430 120]{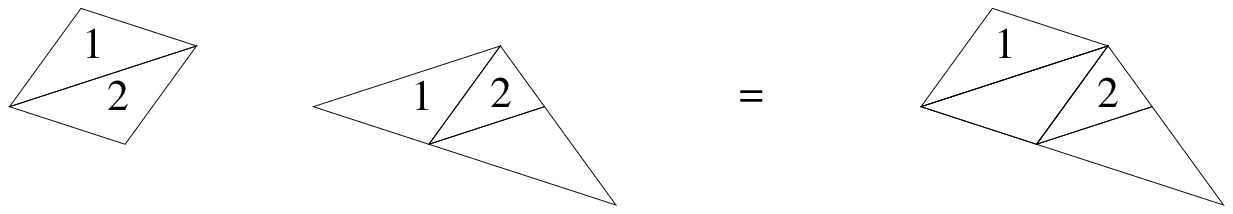}

The inversion is given by
\be
(M_{x_1x_2})^{-1} = M_{x_2x_1}.
\ee
It is straightforward to see that $\einp$ equals the order relation of the
\AG. So we will leave the dot away.

We consider $\mTxx$ as topological \AG\ carrying the discrete topology.
The space of units $\mTx$ is given by the elements of the form $M_{xx}$
which we simply write $M_x$ and call \mix es. The maps $L$ and $R$ act as
\be
L(M_{xy})=M_{xx}\quad\quad R(M_{xy})=M_{yy}.
\ee
Let $\kT$ be the set of \ti s of $T$ where we identify \ti s if they
are mapped onto each other by a translation which is a multiple of a
period of the tiling.
(Since $T$ is an equivalence class we are forced to do this.)
In particular, for tilings which are completely periodic, i.e.\
periodic  in $d$ independent directions, $\kT$ is a finite set.
We call a tiling together with a chosen tile a pointed tiling and denote it
by $T_x$.
Let $M_r(T_x)\in\mTx$ be the smallest \mix\
which covers all balls of radius $r$ having their center inside \ti\ $x$
and keep this tile chosen.
Define $r_c(T_x,T_{x'}):=\sup\{r|M_r(T_x)=M_r(T_{x'})\}$.
Then $d(T_x,T_{x'}):=e^{-r_c(T_x,T_{x'})}$ yields a metric on $\{T_x|x\in\kT\}$
\cite{Ke2}.
\bd
The hull $\Om$ of $T$ is the completion of $\{T_x|x\in\kT\}$ with respect to
the tiling metric $d(T_x,T_{x'})=e^{-r_c(T_x,T_{x'})}$.
\ed
Note that $\Om$ is a finite set if and only if $T$ is completely periodic.
The elements of $\Om$ may be interpreted as \tilx s since any of their
patterns is eventually determined by some $T_x$.
In other words $\om\in\Om$  may be thought of as being
approximated by an increasing chain of \mix es
(which we now write shorter $u$).
\bd
A sequence of \mix es $\{u_\nu\}_\nu$ is called approximating
if
\begin{enumerate}
\item
any finite number of elements is composable, i.e.\
$\forall k\geq 1:u_1\dots u_k \cp u_{k+1}$,
\item
$\rad(u_\nu)\to\infty$ where $\rad(u)$ is the largest $r$ such that all
balls of radius $r$ having their center inside the pointed tile are
covered by $u$.
\end{enumerate}
\ed
An approximating sequence approximates a unique \tilx\ which we write
as its limit, namely
$\om=\lim u_\nu$ is the \tilx\
which contains all pointed patterns $u_\nu$ at its pointed tile.
It is clear that any \tilx\ has an approximation
(just choose a divergent sequence $\{r_\nu\}_\nu$ in $\Real^+$
and take $M_{r_\nu}(\omega)$) so that
we may carry over the notion $\ein$
by writing
$u\ein \omega$ if $u\ein u_\nu$ for some element of an approximation
of $\omega$.

The \miii es of $\om$ are \miii es of $T$ as well, i.e.\
for all $\om\in\Om$, $\mtxx{\om}\subset \mTxx$. One says
that $\omega$ is locally homomorphic to $T_x$ (or $T$).
Consequently $\om$ (or its class) is called locally isomorphic
to $\om'$ if $\mtxx{\om}=\mtxx{\om'}$.
$\tc$ is called minimal if it is locally isomorphic to any $\om\in\Om$
which is equivalent to saying that the closure of the orbit of any $\om$
equals $\Om$.

Since the range of $d(T_x,\cdot)$
is discrete away from $0$ the hull is completely disconnected
(zero dimensional). This means that the topology is generated
by sets which are both open and closed. In fact,
to any unit $u\in\mTx$
we may assign an open and closed subset
\be
U_u:=\{\omega\in\Om|u\ein \omega\}
\ee
of $\Om$. The collection of all of these yields a basis for the topology.

Let $\mTzw$ denote the set of all \mixx es which are composed of two
neighbored tiles (i.e.\ which have boundaries with nonzero intersection)
and such that the first pointed tile is not equal to the second.
$\mTzw$ generates $\mTxx\backslash \mTei$ where $\mTei$ are the \mix es
consisting of one tile only.
The following compactness condition will be required furtheron.
\bi
\item
$\mTzw$ is a finite set.
\ei
Under this condition $\Om$ is compact. In fact,
this condition is equivalent to the requirement that
for any $r>0$ there are only finitely many
 \mix es which are covered by a ball of radius $r$,
and the latter has been used
to prove compactness of $\Om$ in \cite{Ke2}.

To keep contact with \cite{Ke2} where proofs of the above statements may be
found, we visualize a (non periodic) pointed tiling $T_x$
as a particular representative of $T$ in $\Real^d$. For this we
choose a reference point $0$ in $\Real^d$ and introduce a puncture for each
\miii\ of a tile (i.e.\ the puncture is a point of the tile)
so that $\tc_x$ is the representative for which \ti\
$x$ lies with its puncture on $0$.\bs

The \AG\ $\mTxx$ acts on $\Om$ from the right (and analogously from the
left). Let
\be
\Om^\ac=\{(\omega,c)|L(c)\ein \omega\}\subset \Om\times\mTxx
\ee
with relative topology. Define $\gamma:\Om^\ac\rightarrow \Om:\,
(\omega,c)\mapsto\omega\act c$ through an approximating sequence:
Let $\lim u_\nu = \omega$ and set
\be
(\lim u_\nu)\act c := \lim R(u_\nu c).
\ee
More pictorially with $\omega=T'_x$ and $M_y\ein T'_x$
 we have $T'_x\act M_{yy'}:=T'_{x'}$
where $x'$ is the \ti\ of $T'$ which coincides with $y'$ once $M_y$ has been
realized as the \mi\ at $x$ in $T'$.
Note that
\be \label{24071}
L(c)\ein \omega \mbox{ and } L(c')\ein \omega\cdot c \quad\iff \quad
c\cp c'\mbox{ and } L(cc')\ein \omega
\ee
and that
in this case
\be\label{24072}
\gamma(\gamma(\omega,c),c')=\gamma(\omega,c\mul c').
\ee
Moreover
$\gamma(\cdot,M_{xy}):U_{M_x}\rightarrow U_{M_y}$ is continuous
so that we might call it a continuous right action of an \AG.
Let us define an equivalence relation on $\Om^\ac$ using approximations
for the elements of $\Om$:
\be
(\lim u_\nu,c)\sim (\lim u_\nu,c')\quad\mbox{whenever}\quad \exists n:
u_n c = u_n c'.
\ee
It is straightforward to see that this definition is independent of the
approximation and that the relation is transitive. We denote
the equivalence classes by $[\omega,c]$. These classes may be visualized
as \tilxx s, namely $[T'_x,M_{yz}]$ can be represented by $T'_{xz}$, $z$ being
the \ti\ of $T'_x$ corresponding to the $z$ in $M_{yz}$ once the latter has
been identified in $T'_x$ such that $y$ coincides with $x$.
\bl\label{l12041}
Let $\Gr$ be quotient of $\Omega^\ac$ by the above equivalence relation
with quotient topology and consider the groupoid structure
defined by
\be\label{06041}
[\omega,c][\omega', c']=[\omega, c\mul c']
\quad\mbox{provided}\quad
\omega'=\omega\act c
\ee
and
\be
[\omega,c]^{-1}=[\omega\act c,c^{-1}].
\ee
Then $\Gr$ is an $r$-disrete groupoid. Its space of units is homeomorphic
to $\Om$.
\el
{\em Proof:}
First, note that (\ref{06041}) is well defined because of
(\ref{24071},\ref{24072}).
It is straightforward to check that $\Gr$ satisfies the axioms of
a groupoid.
The topology of $\Gr$ is generated by sets of the form
\be
\U_c:=\left[U_{L(c)}\times\{c\}\right]=\{[\omega,c]|L(c)\ein\omega\}.
\ee
$\U_c^{-1}=\U_{c^{-1}}$ shows continuity of the inversion and
$m^{-1}(\U_c)=\bigcup_{c_1\cp c_2, c\ein c_1c_2}\U_{c_1}\times\U_{c_2}$
that of multiplication $m:\Gr^\cp\to\Gr$.
The space of units is $\{[\omega,u]|u=L(u)\ein\omega\}$ which is
open and homeomorphic to $\Om$ via $[\omega,u]\mapsto \omega$.\eb
\bd
We call the groupoid $\Gr$ of Lemma~\ref{l12041} the groupoid associated to
$T$.
\ed
The orbit of $\Gr$ through $\omega$ is given by
$[\omega]_o=\{\omega\act c|L(c)\ein \omega\}$ or more
pictorial $[T'_x]_o=\{T'_x|x\in\kt{T'}\}$.
Hence $[T'_x]_o$ is the equivalence class under translation $T'$:
$T_x \sim_o T'_y$ whenever $T=T'$.
If $T$ is not periodic then the quotient topology on the
space of orbits $[\Om]_o$ is not Hausdorff. Such a
space is an example of a non commutative space in the sense of Connes
\cite{Cone}.
But the reader should be warned that what is called
the non commutative space of
Penrose tilings in \cite{Cone} does not agree with the space of orbits
of the hull of a Penrose tiling but is rather a quotient of it.
We shall comment on that in section~\ref{03081}.

Let $\tilde{\gamma}:\Gr\rightarrow\Om\times\Om$ be given by
$\tilde{\gamma}[\omega,c]=(\omega,\omega\act c)$.
\bd The principal groupoid $\Gs$ associated to $T$
is the image of $\Gr$ under $\tilde{\gamma}$ with weak topology
and groupoid structure defined by the orbit equivalence under $\Gr$.
\ed
\bl
$\Gs$ is isomorphic to $\Gr$ if and only if $T$ is non periodic.
\el
{\em Proof:} Clearly $\tilde{\gamma}$ is a surjective and open
homomorphism of topological groupoids.
But it is only injective if $\gamma$ is free in the sense that
$\omega\act c=\omega$ implies that $c$ is a unit. This is precisely the
case if $T$ is not invariant under a translation.\eb\bs

In particular $\Gr$ is principal if $T$ has no periodicity.
Note that the topology of $\Gs$ does not coincide with the relative topology
inherited from $\Om\times\Om$. Its topology is generated by
\be
U_{M_{xy}}:=\tilde{\gamma}[\U_{M_{xy}}]=\{(T'_{x'},T'_{y'})|M_x\ein T'_{x'},
T'_{y'}=T'_{x'}\act M_{xy}\}.
\ee
In \cite{Ke2} the principal groupoid of a tiling was used instead of
$\Gr$ but it turned out that from the point of view of physics
the groupoid-\CA\ of $\Gs$ is not appropriate for tilings
which have a periodicity in that it does not contain
all translation operators in that case.

\subsubsection{The integer group of coinvariants associated to a tiling}

An important ingredient for the characterization of a tiling is the
integer group of coinvariants of its associated groupoid.
This notion is borrowed from the case of transformation groups
$X\times_\alpha S$ where the action of $S$ on $X$ gives rise to an
action on the group $C(X,\Z)$ of continuous integer-valued functions,
$\alpha(s)^*(f)(x)=f(x\cdot s )$. The coinvariants are then
the quotient of $C(X,\Z)$ by the subgroup generated by
$\{f-\alpha(s)^*(f)|f\in C(X,\Z),s\in S\}$.\bs

The ample semi group $\asg(\Gru)$ of an $r$-discrete groupoid
$\Gru$  is the sub-inverse-semi-group of
$\isg(\Gru)$ given by its compact open $\Gru$-sets.
$\asg(\Gru)$ acts on $\asg(\Gru)^0$ by conjugation,
$U\cdot \U := \U^{-1} U\U$,
leading to a partially defined action on $\Gru^0$ \cite{Ren}:
for $u\in L(\U)$,
\be
\{u\}\cdot \U=\U^{-1}\{u\}\,\U=\{c^{-1}uc\}
\ee
where $c=L^{-1}(u)$. The integer group of coinvariants of
$\Gru$
(or group of coinvariants with coefficients in $\Z$)
shall be the coinvariants of the $\Z$-module
 $C_c(\Gru^0,\Z)$
of integer valued continuous functions over $\Gru^0$ with
compact support
 with respect to the above partially
defined action of $\asg(\Gru)$. More precisely define
$\eta:\asg(\Gru)\to C_c(\Gru^0,\Z)$ by
\be \label{23061}
\eta(\U) := \Ch_{L(\U)}-\Ch_{R(\U)}
\ee
where $\Ch_{U}$ is the characteristic function on $U$ and
let $E_\Gamma$ be the subgroup of $C_c(\Gru^0,\Z)$ generated by $\im\eta$.
\bd
The integer group of coinvariants of an $r$-discrete groupoid $\Gru$ is
$$H(\Gru)=C_c(\Gru^0,\Z)/E_\Gamma.$$
\ed
If $\U$, $\U'$ are disjoint $\Gru$-sets and $\U\cup\U'$ is a $\Gru$-set
as well then $\eta(\U\cup \U') = \eta(\U) +\eta(\U')$.

For transformation groups $\asg(X\times_\alpha S)$ contains
disjoint unions of sets of the form
$U\times\{s\}$ where $s\in S$ and $U\subset X$ is open and compact.
Since
$\eta(U\times\{s\}) = \Ch_U-\Ch_{U\cdot s}$ 
the above definition is indeed a generalization.

Let us consider groupoids associated to tilings.
Since the restrictions of $L$ and of $R$ to the sets $\U_c$ generating the
topology
are both injective $\asg(\Gr)$ contains unions of these sets.
These unions may be chosen to be disjoined because the
hull is zero dimensional.
Now using $L(\U_c)=\U_{L(c)}$ one obtains
\be
H(\Gr) = \langle\{\Ch_{U_u}|u\in \mTx\}\rangle/
\langle\{\Ch_{U_{L(c)}}-\Ch_{U_{R(c)}}|c\in \mTxx\}\rangle.
\ee
We call $H(\Gr)$ also the integer group of coinvariants associated to $\tc$.

\subsection{Reduction}

The concept of reduction is very important in the sequel.
In the context of dynamical systems it is sometimes called induction
but we prefer the notion established for groupoids \cite{Ram}.\bs

Let $\N$ be a sub-\AG\ of $\mTxx$ and define
\be
\Om_\N := \bigcup_{c\in\N} U_{R(c)}
\ee
as well as
\be
\Om^\ac_\N := \Om^\ac\cap(\Om_\N\times \N)
\ee
with relative topologies.
We shall often consider the case in which $\N$ is finitely generated,
i.e.\ $\N= \langle\C\rangle$ for
a finite subset $\C$ of $\mTxx$, then
$\Om_\N = \bigcup_{c\in\C} (U_{R(c)}\cup U_{L(c)})$.
In fact,
since $\Om$ is compact and  the $U_{u}$ form a basis of its topology
any open and closed subset of $\Om$ arrises in this manner with a
finitely generated $\N$.
There are two groupoids which naturally arrise from such a situation, one
is the so-called reduction of $\Gr$ by $\Om_\N$:
\be
\Gr_{\Om_\N}:=\{[\omega,c]\in\Gr|\,\omega,\omega\act c\in\Om_\N\}
\ee
(with relative topology), and the other is
\be
\Gr_\N:=\left[\Om_\N^\ac\right]_\N
\ee
where $[\omega,c]_\N=\{(\omega,c')\in [\omega,c]|c'\in\N\}$
(with quotient topology).
\bd
We call $\N$
approximating if any $\omega\in\Om_\N$ is approximated by a
sequence $\{u_\nu\}_\nu$ with elements in $\N$ and generating if
$[\omega,c]_\N\mapsto [\omega,c]:\,\Gr_\N\to\Gr_{\Om_\N}$ is an
isomorphism.
We call $\N$ regular
if any orbit of $\Gr$ intersects $\Om_\N$, i.e.\ if
$[\Gr]_o=[\Gr_\N]_o$.
\ed
Note that $[\omega,c]_\N\mapsto [\omega,c]:\,\Gr_\N\to\Gr_{\Om_\N}$
provides an isomorphism if it is surjective.
Note also that for minimal $T$ any non-empty $\C\in\mTxx$
generates a regular sub-\AG\ $\erzC$.
None of the properties of the definition implies
any other as counterexamples show. However,
if $\N$ is generating then
$\tilde{\N}=\{c_1uc_2|c_i\in\N,c_1\cp c_2,R(c_1),L(c_2)\ein u\in\mTx\}$
is approximating and $\Gr_\N=\Gr_{\tilde{\N}}$.
\bl
If $\N$ is approximating then the
topology of $\Om_\N$ is generated by sets of the form
$U_u$ with $u\in\N^0$.
\el
{\em Proof:}
It is clear that the topology of $\Om_\N$ is generated by sets of the form
$U_{uv}$ where $u\in \N^0$ and $u\cp v\in\mTx$. If $\N$ is approximating
then there exists for any $\omega\in U_{uv}$ a $u_\omega\in\N^0$
such that $uv\ein u_\omega\ein\omega$. Hence $U_{uv}=\bigcup_{uv\ein\omega}
U_{u_\omega}$. 
\eb
\bl\label{18071}
If $\N$ is regular then
 for any $u\in\mTx$ there is a finite subset $\{c_i\}_i\subset\mTxx$
with $U_u=\dot{\bigcup}_i U_{L(c_i)}$ and $U_{R(c_i)}\subset\Om_\N$.
\el
{\em Proof:}
By regularity we may find for $\omega\in U_u$ a $c\in\mTxx$ such
that $\omega\cdot c\in\Omega_\N$.
Let
$r=\sup_{\om\in U_u}
\inf\{r'|u\ein M_{r'}(\om),\exists c\in\mTxx:\om\cdot c\in\Om_\N,L(c)\ein
M_{r'}(\om)\}$.
Since $U_u$ is compact $r$ is finite.
Choose for any $\om\in U_u$ a $c_\om$ so
that $\om\cdot c_\om\in\Om_\N$ and $L(c_\om)=M_r(\om)$.
In particular $U_{R(c_\omega)}\subset\Om_\N$ and
$\omega\in U_{L(c_\omega)}\subset U_u$.
Hence $U_u=\bigcup_{u\ein\omega}U_{L(c_\omega)}$. Since
$U_{M_r(\om)}\cap U_{M_r(\om')}$ is either $U_{M_r(\om)}$ or empty
the union can be made disjoint and then, by compactness, finite. \eb\bs

The
embedding $i:\Gr_{\Om_\N}\hookrightarrow \Gr$ furnishes a
surjective group homomorphism $i^*:C(\Om,\Z)\to C(\Om_\N,\Z)$ which
satisfies $i^*\circ\eta=\eta\circ i^{-1}$. Therefore
$E_{\Gr_{\Om_\N}}=E_\Gr\cap C(\Om_\N,\Z)$ and $i^*$ induces a surjective
homomorphism $H(\Gr)\to H(\Gr_{\Om_\N})$.
\bl \label{23051}
If $\N$ is regular then
$H(\Gr)=H(\Gr_{\Om_\N})$.
\el
{\em Proof:}
We have to show that
that for $u\in\mTx$ there is a finite $\{u_i\}_i\subset\mTx$ with $U_{u_i}
\subset\Om_\N$ and $[\Ch_u]_{E_\Gr}=\sum_i [\Ch_{u_i}]_{E_{\Gr}}$.
This follows immediately from the foregoing lemma taking $u_i=R(c_i)$. \eb\bs

{\em Remark:} Lemma~\ref{18071} implies even that the groupoids
$\Gr$ and $\Gr_{\Om_\N}$ are continuously similar: Let $\C$ be the
collection of $c_i$'s such that $\Om=\dot{\bigcup}_i U_{L(c_i)}$ and
$U_{R(c_i)}\subset\Om_\N$. Define $\Om\to\C:\om\mapsto c_\om$,
$c_\om$ being the unique element satisfying $L(c_\om)\ein \om$, and
$\epsilon:\Gr\to\Gr_{\Om_\N}$:
\be
\epsilon[\om,c]=[\om\cdot c_\om,c_\om^{-1}cc_{\om\cdot c}].
\ee
Clearly $\epsilon$ is continuous and
$\epsilon\circ i = id_{\Gr_{\Om_\N}}$
for the embedding $i:\Gr_{\Om_\N}\to \Gr$.
But also
 $i\circ\epsilon$ is similar to
the identity on $\Gr$, namely the map $\Theta:\Om\to\Gr$:
$\Theta(\omega)=[\omega,c_\om]$ satifies
$\Theta(L([\om,c]))\epsilon([\om,c])=[\om,c]\Theta(R([\om,c]))$.
This is a continuous version of the theorem of Ramsay \cite{Ram}
guaranteeing that the continuous cohomology of $\Gr_{\Om_\N}$ is
isomorphic to that of $\Gr$ \cite{Ren}.

\subsection{Locally defined maps between tilings}\label{11071}

We want to formulate in the present framework
the concept of locally deriving one tiling from another,
a concept which has been introduced for tilings in \cite{BSJ}.
Roughly speaking a local derivation amounts to a deformation of \miii es,
e.g.\ by adding or deleting or deforming boundaries or decorations of tiles.
Such a deformation will be given by a certain class of maps from
sub-\AG s of $\mTxx$ to the \mixx es $\mTxx'=\mtxx{T'}$ of the other
tiling class $T'$. The results will be partly needed in the third section.

\bl\label{27061}
Let $\N$ be an approximating sub-\AG\ of $\mTxx$ and
$\svarphi:\N\to\mTxx'$ be a map which preserves composability,
commutes with the inverse map 
and satisfies
\bi
\item[1]
$\svarphi(c)\svarphi(c') \ein \svarphi(cc')$
\item[2]  
(growth condition)
there is a $t>0$ such that $|\rad(\svarphi(u))-t\,\rad(u)|$ is
a bounded function on $\mTx$.
\ei
Then
\be
\varphi[\lim u_\nu,c]_\N := [\lim\svarphi(u_\nu),\svarphi(c)]
\ee
defines a homomorphism $\varphi:\Gr_\N\rightarrow\Gr'$.
\el
{\em Proof:}
Let us first show that $\sva$ preserves the order. $c\ein c'$ is by
definition $c\cp {c'}^{-1}$ and $c'=cR(c')$. But then
$\sva(c)\ein \sva(c)\sva(R(c'))\ein \sva(c')$.
Now let $\{u_\nu\}_\nu$ be an approximating sequence. Then for any $1<k$,
$u_1\dots u_k\cp u_{k+1}$. The first condition and preservation of
composability immediatly implies that
$\sva(u_1)\dots \sva(u_k)\cp \sva(u_{k+1})$. By the growth condition
the radius of
$\{\sva(u_\nu)\}_\nu$ diverges so it
is an approximating sequence as well. Hence for $\om=\lim u_\nu$,
$\varphi(\omega)=\lim \svarphi(u_\nu)$ is well defined and clearly continuous.
As well is
$\Om_\N\times \N\to \Om'\times\mTxx'$: $(\omega,c)\mapsto
(\varphi(\omega),\svarphi(c))$.
Thus we are left to show that $\varphi$ induces a map on $[\Om_\N^\ac]_\N$
and preserves the groupoid structure.
(By definition of the topology its continuity is then guaranteed.)
We have $\sva(L(c))\succeq L(\sva(c))$ and hence
$\sva(R(u_\nu c))\succeq R(\sva(u_\nu)\sva(c))$. Therefore
does $L(c)\ein\om$ imply  $\sva(L(c))\ein\varphi(\om)$ and
$\varphi(\om\cdot c)=\lim\sva(R(u_\nu c))=\lim R(\sva(u_\nu)\sva(c))=
\varphi(\om)\cdot\sva(c)$.
This shows that
$\varphi(\Om_\N^\ac)\subset{\Om'}^\ac$ and that $\varphi$ maps equivalent
elements onto equivalent ones, because
equivalence may be expressed as
$(\om,c)\sim (\om,c')$ whenever $\exists c'':c,c'\ein c''$ and $L(c'')\ein\om$.
That $\varphi$ commutes with inversion on the level of groupoids follows from
the corresponding hypothesis on the level of the \AG s.
Finally, multiplication is preserved if
$(\varphi(\om),\sva(c)\sva(c'))\sim (\varphi(\om),\sva(cc'))$. In view of the
above characterization of the equivalence this follows from
 $\svarphi(c)\svarphi(c')\ein\svarphi(cc')$. \eb\bs

We say that $\varphi:\Gr_\N\rightarrow\Gr'$ is locally defined.
The corresponding homomorphism on the principal groupoids is simply
$(\varphi,\varphi):\Gs_\N:=\Gs\cap(\Om_\N\times\Om_\N)\to\Gs'$.
Details of the proof of the following theorem, which is
constructive, will be used in section 3.
\bt\label{25071}
Let $\varphi:\Gr\to\Gr'$ be a locally defined homomorphism.
If $\varphi$ has a left inverse then this left inverse is locally defined.
\et
{\em Proof:}
Let $a\in\mTei$, $\mTei$ denoting the set of \mix es of tiles.
Since $\varphi(U_a)$ is open and compact we may find  a finite subset
$\Phi(a)$ of $\mTxx'$ satisfying
\be \label{26061}
\varphi(U_a)=\bigcup_{v\in \Phi(a)}U_{v}.
\ee
Let $E^0(u)=\{e\in\mTxx|R(e)=u\}$ and denote by $x(e)$ the first pointed
tile of $e$.
The expression $u=\prod_{e\in E^0(u)} e^{-1}x(e)e$ motivates to define
\be
\Phi^e(u):=\{\svarphi(e^{-1})\}\Phi(x(e))\{\svarphi(e)\}
\ee
where we have used set multiplication, and
\be
\Phi(c):=\left(\prod_{e\in E^0(L(c))}\Phi^e(L(c))\right)\{\svarphi(c)\}.
\ee
We claim that
\be\label{27062}
\varphi(\U_c) = \bigcup_{d\in\Phi(c)}\U_d.
\ee
Indeed, using that $u\ein\om$ whenever $\forall e\in E^0(u):
x(e)\ein\om\cdot e^{-1}$ we get
\begin{eqnarray*}
\varphi(\U_c) & = & \bigcap_{e\in E^0(L(c))}
\{[\varphi(\om),\svarphi(c)]|\om\cdot e^{-1}\in U_{x(e)}\} \\
& = &
\{[\om',\svarphi(c)]|
\forall e\in E^0(L(c)): \om'\in
\bigcup_{v\in\Phi(x(e))}U_{\svarphi(e^{-1})v\svarphi(e)}\} \\
&= &
\{[\om',\svarphi(c)]|\om'\in
\bigcup_{v\in\Phi(L(c))}U_v\}
\end{eqnarray*}
from which the claim follows. In particular, setting
$\N=\Phi(\mTxx)$, we have $\im\varphi=\Gr'_\N$.
Furthermore let $\Phi(c)\cp \Phi(c')$, which shall mean that
$\exists d\in\Phi(c)\,\exists d'\in\Phi(c'):d\cp d'$. Then from
(\ref{27062}) we may conclude that $U_{R(c)}\cap U_{L(c)}\neq\emptyset$. Hence
\be\label{27063}
 \Phi(c)\cp \Phi(c')\quad\mbox{implies}\quad c\cp c'.
\ee
Now set $W(d)=\{c\in\mTxx|d\in\Phi(c)\}$ which by construction is finite.
If $c,c'\in W(d)$ then by (\ref{27062}) $\U_c\cap\U_{c'}\neq\emptyset$
and hence $c$ and $c'$ have a common greater element, e.g.\ $L(c)c'$.
Since $L(d)\in\Phi(L(c))\cap \Phi(L(c'))$ also $L(d)\in\Phi(L(c)L(c'))$
so that $d\in\Phi(L(c)c')$. It follows that
$\min\{c|\forall c'\in W(d):c'\ein c\}=\max W(d)$. Define
$\spsi:\N\to\mTxx$:
\be
\spsi(d):=\max W(d).
\ee
It commutes with the inverse.
Because of $d\in\Phi(\spsi(d))$ and (\ref{27063})
$d\cp d'$ implies $\spsi(d)\cp \spsi(d')$.
But it also implies $W(d)W(d')\subset W(dd')$
and hence $\spsi(d)\spsi(d')\ein\spsi(dd')$.

Since $\Phi(A)$ is finite $\spsi$ and $\spsi\circ\svarphi$
satisfy the growth condition (with $t^{-1}$ resp.\ $1$ in place of
$t$) and $\N$ is approximating. Hence they extend to a homomorphisms
$\Gr'_\N\to \Gr$ resp.\ $\Gr\to \Gr$.
Since $\psi\circ\varphi$ extends to the identity
$\psi$ extends to a left inverse.\eb\bs

We say that $\varphi$ is locally invertible (on its image).
\bl
If $T$ is non-periodic then the \AG\ $\N=\Phi(\mTxx)$ occurring in the proof
of the above theorem is generating.
\el
{\em Proof:}
Consider
the commuting diagram
\begin{eqnarray*}
\Gr & \stackrel{\varphi}{\leftrightarrow} & \Gr'_{\N}\\
\tilde{\gamma}\updownarrow  & & \downarrow \tilde{\gamma}' \\
\Gs& \stackrel{(\varphi,\varphi)}{\to} &
\Gs'_{\N}
\end{eqnarray*}
where $\tilde{\gamma}$ is an isomorphism since $T$ is non periodic.
Clearly $(\varphi,\varphi)$ is injective. Suppose that it is not
surjective.
Then there must be an $x\sim_o y$ such that
$\varphi^{-1}(x)\not\sim_o \varphi^{-1}(y)$. But this contradicts that
$\varphi^{-1}$ is a left inverse which preserves (like any
homomorphism of groupoids) orbits.
Therefore $\tilde{\gamma}'$ is bijective and hence $\N$ generating.\eb\bs

Counterexamples show that the requirement for $T$ to be non periodic is
neccessary.

\subsection{Decorations of $\Z^d$}

For many tilings the groupoid is a transformation group the group being
$\Z^d$. This is e.g.\ the case
if the tiling is built from unit cubes which nicely touch at the faces.
But since we allow for \mk ions these tilings do not have to be periodic.
They are called decorations of $\Z^d$.
In any case they allow for $d$ commuting  homeomorphisms of the hull
which are given by the $d$ independent shifts such that the orbits of $\Gr$
become orbits under the action of $\Z^d$ and $\Gr$
a transformation group $\Om\times_\alpha\Z^d$.
The structure of the $K$-groups of \CA s defined by such
transformation groups is fairly well understood and it turns out to be
useful as well in the case where $\Gr$ merely has
a reduction $\Gr_{\Om_\N}$ which is isomorphic to $\Om_\N\times_\alpha\Z^d$.
In view of Lemma~\ref{23051} this already implies for regular $\N$ that
the integer group of coinvariants coincide.
\bd
We say that $T$ reduces to a decoration of $\Z^d$ if there is
a regular $\N$ such that $\Gr_{\Om_\N}\cong\Om_\N\times_\alpha\Z^d$.
If $\N=\mTxx$ then $T$ is called a decoration of $\Z^d$.
\ed
Tilings which reduce to
decorations of $\Z^d$ yield by definition $d$-dimensional topological
dynamical systems. The topological space is $\Om_\N$ which is homeomorphic
to the Cantor set. These systems are
topologically transitive and the notion of minimality in the sense of
tilings and in the sense of dynamical systems coincide.\ms

There is a large class of tilings 
which reduce to decorations of $\Z^d$. These
tilings are composed of $d$ dimensional
parallel epipeds which may be decorated.
For their definition consider a set of $N$ 
vectors $\xi_1,\cdots,\xi_N$ which span $\Real^d$.
A set $J\subset \{1,\cdots,N\}$ containing $n$ elements, $n\leq d$, such that
$\{\xi_i\}_{i\in J}$ are linear independent defines a subset
$\{\sum_{i\in J}c_i\xi_i|c_i\in [0,1]\}$ of $\Real^d$ which is an
$n$ dimensional parallel epiped. We will call a translate of it
an $n$-facet of type $J$.
Their boundaries are $n'$-facets,
$n'<n$, of type $J'\subset J$.
Let $\tc$ be a tiling of $\Real^d$ consisting
of possibly \mk ed \pe{d}s which are arranged
in such a way that
\bi
\item[D1] the boundaries of \ti s overlap, if at all, at common complete
$d'$-facets, $d'<d$.
\ei
Provided the
number of different \mk ions is finite, which we assume,
there is only a finite number of \miii es consisting
of two tiles which touch each other so that the hull is compact.

If $d=N$ then $\tc$
itself a decoration of $\Z^d$ so here we are
interested in $d<N$.
Now fix a type $I_0$ of some \ti\ appearing in $T$ and set
\be\label{19061}
\C=\{a\in\mTei|a\mbox{ is of type }I_0\}.
\ee
where $\mTei$ denotes the set of \mix es of tiles.
For any $i\in I_0$ choose a (common) normal for $d\!-\!1$-facets of type
${I_0\backslash\{i\}}$, i.e.\ a positive and a negative side,
and let $\B_i\subset\mTzw$ consist of those \mixx es which have as
common boundary a \pe{d-1}\ of type ${I_0\backslash\{i\}}$ and such that
their first resp.\ second pointed tile is on the positive resp.\ negative
side of the common boundary. Then $\Om_{\erzC}=\bigcap_i\Om_{\erz{\B_i}}$.
Define $\breve{\alpha}_i:\Om_{\erz{\B_i}}\to\Om_{\erz{\B_i}}$ by
\be
\breve{\alpha}_i(\omega) = \omega\cdot c
\ee
where $c$ is the unique element of $\B_i$ for which $L(c)\ein\omega$.
Clearly $\breve{\alpha}^{-1}_i(\omega) = \omega\cdot c^{-1}$
with the unique $c$ for which $R(c)\ein\omega$.
Now we require in addition that
\bi
\item[D2]
for all $i\in I_0$ and all $\om\in \Om_{\erzC}$ there are natural numbers
$n,m>0$ such that $\breve{\alpha}_i^n(\om)\in \Om_{\erzC}$ and
$\breve{\alpha}_i^{-m}(\om)\in \Om_{\erzC}$.
\ei
Let $n^\pm_i(\om)$ be the smallest $n>0$
for which $\breve{\alpha}_i^{\pm n}(\om)\in \Om_{\erzC}$.
This defines maps $n^\pm_i:\Om_{\erzC}\to\Z^{>0}$ which
are continuous by the compactness of $\Om_{\erzC}$.
Consequently $\alpha:\Om_{\erzC}\to\Om_{\erzC}$
\be
\alpha_i(\om):=\breve{\alpha}_i^{n^+_{i}(\om)}(\om)
\ee
are continuous maps (first return maps) having inverses
$\alpha_i^{-1}:=\breve{\alpha}_i^{n^-_{i}}$.
\bl
Let $\tc$ be a $d$ dimensional
tiling which is composed of (possibly \mk ed) parallel epipeds
and satisfies conditions {\rm D1,D2}.
Then the homeomorphisms $\alpha_i$ defined above commute pairwise.
Moreover, $\erzC$ defined by (\ref{19061}) is regular.
\el
{\em Proof:}
Draw in any \ti\ which has \pe{d-1}s of type
${I_0\backslash\{i\}}$ a line segment
joining the middle points of the boundary facets of
this type. (The middle point of an \pe{n}\ of type $J$ is at
$\sum_{i\in J}\frac{1}{2}\xi_i$.) This segment goes through the
middle point of the tile.
 All the line segments in a tiling fit together to yield a set of infinite
\faden s in $\Real^d$.
Such a \faden\  will be called $i$-\faden\ its type being $i$.
As an example the below figure
shows a patch of an octogonal tiling with its line segments of two types.
The four vectors $\{\xi_1,\dots,\xi_4\}$ are given to the right,
the \faden s are of type $1$ and $3$.

\epsffile[0 0 420 260]{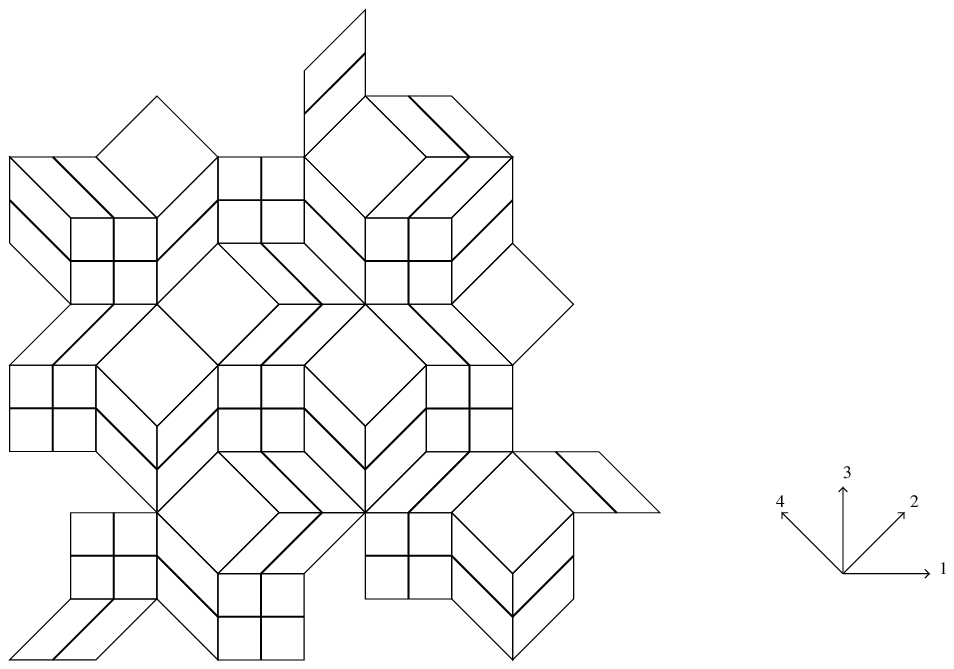}

In general the \faden s have the properties:
\bi
\item[1]
If $l\neq l'$ 
then $l\cap l'$ contains at
most one point (which is then the middle point of the tile through which
they both go),
 and if $l$ and $l'$ are of the same type then  $l\cap l'=\emptyset$.
\item[2]
Let $\pi_i$ be the orthogonal projection along the span of $\{\xi_j\}_{j\in
I_0\backslash\{i\}}$.
Then, for $d=2$ and any \faden\  which is of type $i$, $\pi_i(l)$
equals to the span of $\xi_i$.
\ei
Any type of \faden\ shall now be given the direction of the chosen
normal of $I_0\backslash\{i\}$.
The \ti s
belonging to it can be ordered and the application of
$\breve{\alpha}_i$ can be geometrically interpreted as a shift
to the next tile along an $i$-\faden .
Condition D2 then reads that any \faden\  of type $i\in I_0$
contains in both directions
infinitely many \ti s of type $I_0$.
Let us single out two elements of $I_0$ which we
denote by $1$ and $2$.
Let $l_0,m_0$ be the \faden s of type $1$ resp.\ $2$ which go through
$x$, the pointed tile of $T_x\in\Om_{\erzC}$. Let $m_1,l_1,m'$ be the
successor resp.\ of $m_0$ along $l_0$, of $l_0$ along $m_0$, and of
$m_0$ along $l_1$, c.f.\ the below figure where this
situation is indicated topologically.

\epsffile[0 0 420 180]{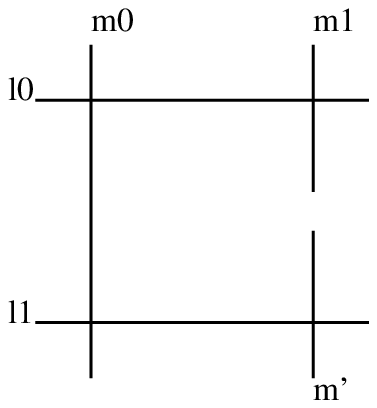}

Then $\alpha_1\circ\alpha_2(T_x)=\alpha_2\circ\alpha_1(T_x)$ if and only if
$m'=m_1$.

We first proof that this is the case for $d=2$.
By  property 1 above
$m'$ can neither intersect $l_1$ at a second
point nor $m_0$ at all
nor can it intersect $l_0$ at a lower point than $m_1$ does.
The same reasoning applies to $m_1$ so that, since $\pi_2(m')$ is the span
of $\xi_2$,
$m'\cap m\neq\emptyset$.
Hence $m'=m_1$. 

The case $d>2$ can be traced back to $d=2$ reducing $T_x$ to a
two dimensional tiling $\overline{T_x}$ by the following process.
Let $J_0=I_0\backslash\{1,2\}$.
We define the $J_0$-neighborhood of a tile $a$ which has
\pe{d-2}s of type $J_0$ to consist of this tile an those tiles
whose boundary intersects the boundary of $a$ at a \pe{d-2}\ of type $J_0$.
Let $\overline{T_x}$
be the smallest connected component consisting of tiles having
\pe{d-2}s of type $J_0$ in such a way that, first, it contains with a tile
its $J_0$-neighborhood, and second, it contains the pointed tile of $T_x$.
In particular $\overline{T_x}$ is a connected topological manifold
which contains complete $1$- and $2$-\faden s.
Let $\pi$ be the orthogonal projection along
$\langle\{\xi_i\}_{i\in J_0}\rangle$. Then $\pi(\overline{T_x})$ is a
two dimenional tiling to which we can apply the above arguement leading
to $\pi(m_1)=\pi(m')$.
To see that this implies $m_1=m'$ consider the homotopy
$F_t(\xi_i)=(1-t)\xi_i+t\pi(\xi_i)$ for $i\notin J_0$ and $F_t(\xi_i)=\xi_i$
for
$i\in J_0$. $F_1(\overline{T_x})$ may be understood as the
Cartesian product of $\pi(\overline{T_x})$ with a $d\!-\!2$-facet of type
$J_0$.
Certainly no new intersections between lines
in $F_t(\overline{T_x})$ occurr for $t>0$ so that
the decomposition shows that $\pi(m_1)=\pi(m')$ whenever $m_1=m'$.

By compactness of $\Om_{\erzC}$ the distance of two neighbored \faden s has an
upper and a lower bound. Hence for some $r$ any $u\in\mTx$ of radius larger
than $r$ contains a
tile of type $I_0$. Therefore $\erzC$ is regular.
\eb

\bt
Let $\tc$ be a $d$ dimensional
tiling which is composed of (possibly \mk ed) parallel epipeds
and satisfies conditions {\rm D1,D2}. Then it reduces to a $Z^d$ decoration.
\et
{\em Proof:}
The last lemma implies that the map
$\Om_{\erzC}\times_\alpha \Z^d\to\Gr_{\Om_{\erzC}}$: $(\omega,e_i)\mapsto
[\omega, c]$
where $c$ is uniquely determined by
$c=c_1\dots c_{n^+(\omega)}$ with $c_k\in\B_i$, $c_k\cp c_{k+1}$,
and $L(c)\ein\omega$, is
an isomorphism of groupoids.\eb\bs

All tilings which are obtained
by the so-called generalized
dual (or grid) method \cite{SoSt2} fall under the vicinity of the above theorem
provided they satisfy D2.
In fact, the lines of different type can be understood objects dual
to the hyperplanes of the grids so that we just inverted the construction of
\cite{SoSt2}.
In particular all tilings obtained by the
cut and projection method \cite{DuKa}
(which is equivalent to the
grid method using special grids only)
reduce to decorations as they all satisfy D2.

Among the \sst\ tilings which reduce to decorations of $\Z^2$
are the Ammann-Beenker-
the Socolar- and the Penrose tilings.

\section{\CA ic characterization of tilings}

We review the \CA ic constructions neccessary to formulate the $K$-theoretical
gap labelling for particles moving in a tiling.
As already mentionned the
$K$-theoretical gap labelling may be seen as an invariant of a \CA.
The natural candidate for this algebra is the algebra $A_\tc$ associated to
the tiling $\tc$ in which the particle is moving since it
contains all local operators involving translations and multiplications
with pattern dependent functions.
\bd
The algebra $\A_\tc$ associated to a tiling $\tc$ is the reduced groupoid-\CA\
$C^*_{red}(\Gr)$ of the groupoid $\Gr$ associated to $\tc$.
\ed
Any $r$-dicrete groupoid defines a reduced groupoid-\CA\ \cite{Ren}.
Specified to the groupoid $\Gr$ it is defined as follows:
Let $C_c(\Gr)$ be the $*$-algebra  of continuous
functions $f:\Gr\rightarrow \Complex$
with compact support and multiplication and involution given by
\begin{eqnarray}                             \label{25061}
f*g\,[\om,c] & = &
\sum_{[\om\cdot c,{c'}]\atop\omega,c\: \mbox{\tiny fixed}}
f[\om,cc']\: g[\om\cdot cc',{c'}^{-1}] , \\
f^*\,[\om,c] & = & \overline{f[\om\cdot c,c^{-1}]}        \label{25062}.
\end{eqnarray}
$C_c(\Gr)$ is generated by the characteristic functions
$e_c$ onto $\U_c$ which satisfy the relations
\begin{eqnarray}
e_c*e_{c'} &=& \left\{
\begin{array}{ll}
e_{c c'} & \mbox{if $c\cp c'$} \\
0 & \mbox{else}
\end{array} \right.\\
e_c^*&=&e_{c^{-1}}.
\end{eqnarray}
Hence in any representation $e_c$ is represented as a partial isometry.
$C^*_{red}(\Gr)$ is the closure of $C_c(\Gr)$
taken with respect to the reduced norm.
This norm is defined using the family of representations $\pi_{[\omega]_o}$,
one for each point in the \nc\ space,
acting on the Hilbert space of square summable functions
$\Psi:[\omega]_o\rightarrow \Complex$ through
\be
(\pi_{[\omega]_o}(e_{c})\Psi)(\omega)=\left\{
\begin{array}{ll}
\Psi(\omega\cdot c) & \mbox{if $L(c)\ein\omega$} \\
0 & \mbox{else}
\end{array} \right..
\ee
Now the reduced norm is given by
 $\|f\|_{red}=\sup_{[\om]_o\in[\Om]_o}\|\pi_{[\om]_o}(f)\|$
where $\|\pi_{[\om]_o}(f)\|$ is the operator norm.
Since the topology of $\Gr$ has a countable basis
$\A_\tc$ is separable.\bs

As an example consider a tiling which is a decoration of $\Z^d$.
That is there are $d$ commuting homeomorphisms $\alpha_i$ of the
hull such that $\Gr=\Om\times_\alpha\Z^d$.
To any continuous 
$f:\Om\times_\alpha\Z^d\to \Complex$ with compact support one may assign
the function $\hat{f}:\Z^d\to C(\Omega)$ through
$\hat{f}(k)(\om) = f(\om,k)$. Carried over from   (\ref{25061},\ref{25062})
multiplication and involution then become convolution resp.\ involution twisted
by
$\alpha$:
\begin{eqnarray}
\hat{f}*\hat{g}\:(k)& =& \sum_{m\in\Z^d} \hat{f}(m)\,
(\hat{g}(k\!-\!m)\circ\alpha(m)) \\
\hat{f}^*(k)& =& \overline{\hat{f}(-k)\circ\alpha(k)}.
\end{eqnarray}
The closure $C^*_{red}(\Om\times_\alpha\Z^d)$ is
isomorphic to $C(\Om)\times_\alpha\Z^d$,
the crossed product of $C(\Om)$ with $\Z^d$ by the action
$\alpha(k)(\hat{f}(m))=\hat{f}(m)\circ\alpha(k)$,
$\alpha(k)=\alpha_1^{k_1}\circ\dots \circ \alpha_d^{k_d}$. 
\bs

Remember that $\tc$ is minimal if the closure of the orbit of any $\om$
equals $\Om$.
The lattice of closed (twosided) ideals of $\A_\tc$
may be identified with the lattice of
open invariant subsets of $\Om$, i.e.\ those open subsets
which contain next to an element $T$ all its equivalent elements \cite{Ren}.
The groupoid $\Gr$ is called minimal if $\Om$ does not contain
any proper invariant open subset. But this is just excluded for minimal
tilings. So
if $\tc$ is minimal then $\Gr$ is minimal
 which implies that $\A_\tc$ is simple \cite{Ren}.
\bs

Lemma~\ref{23051} has a \CA ic counterpart.
Two \CA s $\A$ and $\B$ are called stably isomorphic if
$\A\otimes\K$ is isomorphic to $\B\otimes\K$, where $\K$ is the algebra of
compact operators (on an infinite dimensional separable Hilbert space).
A theorem  of Brown \cite{Bro} states
that a full reduction $\B$ of $\A$ is \si\ to $\A$.
In this context a reduction of $\A$ is determined by a projection $p\in\A$,
namely given by $\A_p:=\{x\in\A|px=xp=x\}$,
and $\A_p$ is full if the twosided ideal
generated by $p$ is dense in $\A$.
\bl \label{22061}
Let $\N\subset\mTxx$ be regular. Then $C^*_{red}(\Gr_{\Om_\N})$
is \si\ to $\A_\tc$.
\el
{\em Proof:}
Let $\Ch_{\Om_\N}$ be the projection in $\A_T$ which is the characteristic
function onto $\Om_\N$ and $f\in C_c(\Gr)$.
Then $f*\Ch_{\Om_\N}=\Ch_{\Om_\N}*f=f$ if and only if
$\mbox{supp}f\subset \Gr_{\Om_\N}$.
Hence $C_c(\Gr_{\Om_\N})$ is a dense subalgebra of $(\A_T)_{\Ch_{\Om_\N}}$
and, since the closure of $C_c(\Gr_{\Om_\N})$ with respect to the
norm of $\A_T$ on the one hand and with respect to reduced norm defined
for $\Gr_{\Om_\N}$ on the other coincide, we have
$(\A_T)_{\Ch_{\Om_\N}} = C^*_{red}(\Gr)_{\Ch_{\Om_\N}}$.
To show that this reduction is full we show that
the two sided ideal generated by $\Ch_{\Om_\N}$ contains $1$,
the unit of the algebra,
 if $\N$ is regular. By
Lemma~\ref{18071} there is a finite set $\{c_i\}_i$ satisfying
$\Om=\dot{\bigcup}_i U_{L(c_i)}$ and $U_{R(c_i)}\subset\Om_\N$.
Then $1=\sum_i e_{c_i}e_{c_i^{-1}}=\sum_i e_{c_i}\Ch_{\Om_\N}e_{c_i^{-1}}$
showing that 1 is contained in the ideal.\eb

\subsection{$K$-theoretical gap labelling}

A local selfadjoint operator $H$ describing the motion of a particle
in the tiling $\tc$ is an operator in
the representation $\pi_T$ of $\A_\tc$, namely
\be \label{14061}
H\psi(T_x)=\sum_{x'\in X(\tc)} H_{x,x'}\psi(T_{x'}).
\ee
Locality refers to the requirement that
the matrix element $H_{x,x'}$ depends only on a the \mixx\
of a certain size to which $x,x'$ belong, i.e.\ that $H=\pi_T(h)$ for some
$h\in C_c(\Gr)$.
In particular $h$ may be of the
form $h=-\Delta + \sum_i v_i e_{u_i}$ where $\{u_i\}_i$
is a collection of \mix es and $v_i\in\Real$
and $\Delta$ is the discrete Laplacian. The latter takes
the form $\Delta=\sum_{A\in\mTzw} b_A e_A -\sum_{a\in\mTei} b_a e_a$ where
$b_a\in\Real$ and $b_A^*=b_{A^{-1}}\in\Complex$.

The values of the integrated density of states $N_H(E)$ of $H$  at energies
$E$ lying in a gap of its spectrum serve as labels for the gaps;
they are insensitive to certain perturbations of the operator.
In Bellissard's $K$-theoretic formulation of the
gap labelling \cite{Be4,Be1} these values 
are recognized as elements of
$\tr_*K_0(\A)$ where $\A$ is a \CA\ 
represented on the above Hilbert
space, $H$ an element of its representation, $\tr$
a trace on $\A$, and $\tr_*$ the induced state on its $K_0$-group $K_0(\A)$:
The $K_0$-group of a unital
\CA\ $\A$ is obtained via Grothendieck's construction
from the monoid of projection classes of $\A\otimes\K$, i.e.\
the equivalence classes of projections of $\A\otimes\K$
under $p\sim q$ whenever $\exists u\in\A\otimes\K:
p = uu^*,q = u^*u$ may be added orthogonally, $[p]+[q]=[p\oplus q]$,
yielding a monoid $V(\A)$, and this monoid may be completed to an abelian group
whose elements are classes of pairs under
$([p],[q])\sim ([p'],[q'])$ whenever
$\exists [r]\in V(A):[p]+[q']+[r] = [q]+[p']+[r]$, see \cite{Bla,Mur} for
details.
Any trace of $\A$ extends to a trace on $\A\otimes\K$ and
defines a linear map $\tr_*:K_0(\A)\to\Real$:
$[p]\mapsto \tr(p)$. The gap labelling in the above formulation requires the
equality
\be \label{28071}
N_H(E) = \tr(\chi_{h\leq E})
\ee
$h$ being the element which is represented by $H$ and $\chi_{h\leq E}$
the spectral projection
of $h$ to energies smaller or equal to $E$. This equality involves
validity of Shubin's formula \cite{Be1} by which the trace is
equated to the operator trace per volume in the corresponding representation.
Taking $\A$ to be the algebra associated to the tiling and considering
the above representation the $K$-theoretical gap labelling then reads:
if $E$ lies in a gap
\be
N_H(E)\in 
\tr_*K_0(\A_\tc)\cap[0,1]
\ee
provided (\ref{28071}) holds. In other words $\tr_*K_0(\A_\tc)\cap[0,1]$
is the set of gap labels predicted by $K$-theory.

One  motivation for using $\A_\tc$ is that
it is expected not to yield to many values on the r.h.s.\ so that
if the couplings $v_i$ are strong and diverse enough all elements of
$\tr_*K_0(\A_\tc)$ actually occur as labels for
 open gaps of $H$. In that case,
and if $\tr$ is faithful on $\A_\tc$,
the density of the values of the integrated density of states
on gaps in $[0,1]$
expresses the fact that the continuous part of the
spectrum is a Cantor set.
In fact, conclusions on the nature of the spectrum may partly be drawn
without the need to connect the gap labelling with the values of
the integrated density of states, any faithful trace may be used.
For instance if $\tr$ is faithful and the set of gap labels
${\cal L}_{\tr} (h)=\{\tr(\chi_{h\leq E})|E\notin\sigma(h)\}$ dense in $[0,1]$
the spectrum $\sigma(h)$ cannot contain a proper closed interval, for,
if $[a,b]\in\sigma(h)$, then by faithfulness
$\tr(\chi_{h\leq b})>\tr(\chi_{h\leq a})$ -- here $\tr$ has to be extended
to measurable functions over $\sigma(h)$ -- so that
$[0,1]\backslash{\cal L}_{\tr} (h)$
would contain the open interval $(\tr(\chi_{h\leq a}),\tr(\chi_{h\leq b}))$.
However, up to now there is no $K$-theoretic formulation of a condition for
a \So\ $h$
under which ${\cal L}_{\tr} (h)$ coincides with $\tr_*K_0(\A_\tc)$.
For a $C^*$-algebraic formulation of such a condition for the discrete
magnetic Laplacian on $\Z^2$ see \cite{Sh2}.

Another consequence of the realization of $H$ as an element of a
representation of $A_\tc$ is that for minimal $\tc$
its spectrum has no discrete part, since the spectral projection onto
the eigenspace of a discrete eigenvalue with finite multiplicity
would have to be represented by a compact operator.
In fact, for minimal $\tc$ the algebra $\A_\tc$ is antilimial \cite{Ped},
and hence does not contain compact operators, namely
it is neither limial itself -- its primitive
spectrum contains one single point whereas two different representations
$\pi_{[\om]_o}$ and
$\pi_{[\om']_o}$ are not unitarily equivalent
 -- nor can it
contain a limial ideal.

\subsubsection*{Traces on $\A_\tc$}

Any normalized trace $\tr$ on $\A_\tc$ restricted to a linear functional
$\mu$ on $C(\Om)$ defines a normalized
measure on $\Om$ also denoted by  $\mu$ through
$\mu(f)=\tr(f)=\int f d\mu$. This measure is invariant under the groupoid
in the sense that $\mu(U_{L(c)})=\mu(U_{R(c)})$.
A direct consequence is that the values of $\mu$ on integer valued
continuous functions over $\Om$ lie in the $\Z$-module generated by
the traces of projections of the algebra, and therefore
\be \label{11051}
\mu(C(\Om,\Z))
\subset\tr_*K_0(\A_\tc).
\ee
The existence of a faithful trace implies that
the notion of positive elements of $H(\Gr)=\C(\Om,\Z)/E_\Gr$ as being those
which have a representative which is a positive function turns $H(\Gr)$ into
an ordered group. That is the positive elements $H^+(\Gr)$ satisfy
$H^+(\Gr)+H^+(\Gr)\subset H^+(\Gr)$, $H^+(\Gr)-H^+(\Gr)=H(\Gr)$ and
$H^+(\Gr)\cap -H^+(\Gr)=\{0\}$. In fact,
the existence of a faithful trace rules out that
a nonzero element can be both positive and negative. It being zero on
$E_\Gr$ it moreover defines a state (if normalized) on $H(\Gr)$.
A similar statement holds true for $K_0(\A_\tc)$: The existence of a normalized
faithful trace guarantees that $\A_\tc$ is stably finite and therefore
the usual notion of positive elements of $K_0(\A_\tc)$ as those
which have a representative which is a projection in $\A_\tc\otimes\K$
turns $K_0(\A_\tc)$ into an ordered group (and the trace
induces a state on that group) \cite{Bla}.

Conversely any $\Gr$-invariant normalized
measure $\mu$ on the hull $\Om$ defines a normalized
trace through
\be \label{24061}
\tr(f):=\int_\Omega\mbox{P}(f)\,d\mu
\ee
where $\mbox{P}:\,C^*_{red}(\Gr)\rightarrow C(\Omega)$ is the
restriction map. $\mbox{P}$ is the unique conditional expectation on
$C(\Om)$ and is faithful \cite{Ren}.
Moreover if $\tc$ is minimal every non trivial
invariant measure has to have closed support $\Omega$ so that $\tr$ defined
by (\ref{24061}) is faithful.\bs

One of the goals of this article is the determination of $\tr_*K_0(\A_\tc)$.
The question under which circumstances a given trace satisfies
Shubin's formula and (\ref{28071}) will not be addressed here, but see
 \cite{Be1,BBG,Ke2} for investigations in this directions.

\subsection{$K_0$-groups for tilings}

The $K_0$-group of a
\CA\  depends together with its order structure only on its stable
isomorphism class.
The same holds true for the $K_1$-group which may be understood as
the $K_0$-group of the suspension of $\A$.
For this reason we may apply the known results on crossed products
to obtain the structure of the $K$-theory for tilings which reduce to
decorations. We are not able to present any results on $K$-groups
in case the tiling algebras are not stably isomorphic to crossed
products.\bs

A lot is known about the $K$-groups of crossed products of the form
$C(\Om)\times_\alpha\Z^d$ in particular for zero dimensional $\Om$.
Recently a relation of these $K$-groups with
the group cohomology $H(\Z^d,C(\Om,\Z))$
of $\Z^d$ with coefficients in $C(\Om,\Z)$
was discovered \cite{FoHu}. We will not discuss group cohomology here
but what is important for us is
that the cohomology group of highest nonvanishing degree (which is $d$)
coincides with the group of
coinvariants of $\Om\times_\alpha\Z^d$ and that the $K_0$-group
decomposes into
\be \label{08075}
K_0(C(\Om)\times_\alpha\Z^d)\cong H(\Om\times_\alpha\Z^d)\oplus H'
\ee
where $H'$ is trivial for $d=1$ and equal to $\Z$ in $d=2$.
(In general $H'$ is a direct sum of cohomology groups of degrees
$d-2n$, $0<n\leq \frac{d}{2}$ and $K_1(C(\Om)\times_\alpha\Z^d)$
is a direct sum of cohomology groups of degrees
$d-1-2n$, $0\leq n<\frac{d}{2}$ \cite{FoHu}.)
Up to $d=3$ this result was
obtained before \cite{Els2} in an even more explicit form
in which in particular becomes clear
that
the image of the state on $K_0(C(\Om)\times_\alpha\Z^d)$ induced by a
trace on
$C(\Om)\times_\alpha\Z^d$ satisfies\footnote{
In \cite{Els2} ergodic measures have been used, but ergodicity not essential
for the proof of (\ref{06061}).}
\be \label{06061}
\tr_*K_0(C(\Om)\times_\alpha\Z^d)=\tr_*(H(\Om\times_\alpha\Z^d)) =
\mu(C(\Om,\Z)).
\ee
(\ref{06061}) holds for arbitrary $d$ under the restriction that
$\Om\times_\alpha\Z^d$ splits into a Cartesian product
$\Om\times_\alpha\Z^d=(\Om_1\times_{\alpha_1}\Z)\times
\dots\times(\Om_d\times_{\alpha_d}\Z)$
\cite{Ke2}.

(\ref{08075}) immediately carries over to tilings which reduce
to decorations of $\Z^d$ due to Lemma~\ref{22061}
\be
K(\A_\tc)\cong K(C(\Om_\N)\times_\alpha\Z^d).
\ee
This also concerns the order structure which is however not known for $d>1$.\ms

As for (\ref{06061})
recall that in case the tiling reduces to a decoration
$\A_\N=C(\Om_\N)\times_\alpha\Z^d$ is a subalgebra of $\A_\tc$, i.e.\ there is
an embedding $\imath:\A_\N\rightarrow\A_\tc$.
This embedding induces an isomorphism
$\imath_*$ from $K_0(\A_\N)$ onto $K_0(\A_{\tc})$.
In fact, for separable \CA s, $\A$ being stably isomorphic to $\B$
is equivalent to the existence
of a (strong) Morita equivalence $\A$-$\B$-bimodule
which may be viewed as an element of $K\!K(\A,\B)$ and is a special
case of a $K\!K$-equivalence \cite{Bla,Ska,Cone}.
Any $K\!K$-equivalence between $\A$ and
$\B$ yields an isomophism from $K\!K(\Complex,\A)$ onto
 $K\!K(\Complex,\B)$, namely
by multiplying it from the right, the multiplication being the Kasparov
product.
Translated into $K_0$-groups, $K\!K(\Complex,\A)$ being isomorphic to
$K_0(\A)$, the right multiplication of elements of $K\!K(\Complex,\A_\N)$
with the canonical Morita equivalence $\A_\N$-$\A_\tc$-bimodule,
which as a linear space is
$\A_\N\A_\tc$,
precisely  becomes $\imath_*$.

Now $\tr\circ\imath$ is a trace
on $\A_\N$ which is normalized to
$\tr(\imath(1_{\A_\N}))=\tr(\chi^{}_{\Om_\N})=\mu(\Om_\N)$.
Since the invariant measure
on $\Om_\N$ corresponding to $\tr\circ\imath$ is $\mu|_{\Om_\N}$ we get
\be
\tr_*\circ \imath_*K_0(\A_\N)=\mu(C(\Om_\N,\Z))
\ee
and therefore
\be
\tr_*K_0(\A_{\tc})=\mu(C(\Om_\N,\Z))\subset \mu(C(\Om,\Z)).
\ee
Together with (\ref{11051}) this extends (\ref{06061}) to tilings
which reduce to decorations of $\Z^d$, $d\leq 3$.

$K_0(\A_\N)$ and $K_0(\A_{\tc})$ differ only in their order units
(the images of the units of the algebras in $K_0$).
If one identifies them as above
the order unit of the former is the $K_0$-class of $\chi^{}_{\Om_\N}$.


\section{Substitution tilings}

We have seen that the integer group of coinvariants furnishes part of the
$K$-theory
of the algebra of the tiling. In particular in one and two dimensions it
yields up to order the $K_0$-group. To determine this group we need further
structure which is provided by a locally invertible \sst.\bs

A \sst\ of a tiling may be thought of as a rule according to which the
\ti s of the tiling
are to be replaced by \mi s which fit together to yield a new tiling.
An algebraic way to formulate this is by means of homomorphisms of
the \AG\ $\mTxx$ incorporating thus their local nature.
For the definition recall that $E^0(R(c))=\{e\in\mTxx|R(e)=R(c)\}$ and
that a homomorphism satisfies the growth condition with $t>0$ if
$|\rad(\hat{\rho}(u))-t\,\rad(u)|$ is a bounded function on $\mTx$.
Let $x(c)$ be the first pointed tile of the \mixx\ $c$.
\bd
A \sst\ of $T$ is a homomorphism of \AG s
\be
\srho:\mTxx\to \mTxx
\ee
which satisfies the growth condition with $t>1$,
and for which the \miii\ of
$\srho(c)$ is composed without overlap (up to boundaries) of
the \miii es of the $\hat{\rho}(x(e))$, $e\in E^0(R(c))$,
and the first resp.\ second pointed
tile of $\srho(c)$ correspond to
the pointed tile of $\hat{\rho}(x(c))$ resp.\
$\hat{\rho}(x(c^{-1}))$.
\ed

Since $\srho$ satisfies the growth condition it
defines a homomorphism of $\Gr$ into itself.
We shall call a tiling resp.\ its class
invariant under a \sst\ if $[\rho]_o(T)=T$. We call it a \sst\ tiling if it
allows for a \sst\ such that
$[\rho]_o(T)$ is locally isomorphic to $T$.

Like any homomorphism $\srho$ is determined by its action on the generators
$\mTzw$. But by the above definition this means that $\srho$ may be
given, first, by its image on the \mix es of tiles, $\mTei$,
and second, by the relative
position of these images.

A \sst\ can be iterated and it is
called primitive if for some $n$ and all $a\in \mTei$ the \miii\
$\rho^n(a)$ contains at least one tile of each tile class.\ms

If a tiling allows for a \sst\ which has a right inverse, and this is
equivalent with locally invertible as we have seen, then the \sst\
can be used to compute the coinvariants associated to the tiling class.
We shall carry this out below but first give a brief discussion on
geometric realizations of \sst s.

\subsection{Deflation}

Geometrically we view a pointed tiling as a representative of $T$ in the
Euclidian space $\Real^d$ with fixed origin $0$. To be precise we
choose a point for each tile class (its puncture). Then $T_x$ shall correspond
to the representative of $T$
for which the puncture of the pointed tile coincides with $0$ thereby
getting a bijective correspondance between pointed tilings and representatives
in $\Real^d$ having the puncture of one of its tiles at $0$.
Substitutions appear then as deflations followed by rescaling.
Let $t^{-1}\tc$ denote $T$ rescaled by $t^{-1}$ and $\mT(T)$ the (unpointed)
\miii es of $T$.
A deflation
is given by:
\bi\item[1]
a pattern class $\trho(a)\in\mT(t^{-1}\tc)$ for each tile class $a$,
$t>0$.
\item[2]
a relative position between
$a$ and $\trho(a)$ for each tile class $a$.
\ei
Let $\tl$
be a representative of $\tc$ in $\Real^d$.
In the process of applying the deflation
any tile of class $a$ in $\tl$ is to be replaced by a pattern of class
$\trho(a)$ (its replacement)
at the relative position given above.
To be more precise choose a tile in $\trho(a)$.
Its puncture shall now indicate the
position of a pattern of class $\trho(a)$ in $\Real^d$. Then the
relative position i.e.\ the difference
between the position of the tile and its replacement shall be $x_a\in\Real^d$.
Thus if  a tile of $\tl$ of class $a$ is at $x\in\Real$ then
a pattern of class $\trho(a)$ is in $\trho(\tl)$ at $x-x_a$.
The object $\trho(\tl)$ 
is a composition of patterns of the classes $\trho(a)$, and it is required
that this yields
a tiling 
which is locally isomorphic to $t^{-1}\tc$. In particular no
overlap and no gaps are allowed.

Such a deflation defines a \sst\ in the algebraic sense (as a homomorphism
of $\mTxx$). By the above choice of a tile in $\trho(a)$
the latter may now be understood
as a \mix\ of $t^{-1}\tc$.
We let $\srho(a)$ be $t\trho(a)$ which is $\trho(a)$ scaled by $t$,
the \miii\ of $\srho(M_{x_1x_2})$ be
$t\trho(M)$, and the first resp.\ second pointed tile be the one
corresponding to the
pointed tile of $\srho(x(M_{x_1x_2}))$ resp.\ $\srho(x(M_{x_2x_1}))$.\ms

A deflation is locally invertible
if the determination of whether or not a \mi\ of $\trho(\tl)$
is a replacement of a \ti\ of $\tl$ may be uniquely carried out by
inspection of the larger patterns around that \mi\ up to a given finite size.
In other words, for each $a$
there is a finite set of pointed pattern classes
$\Phi_t(a)\subset\mtx{t^{-1}\tc}$ with
$\forall v\in\Phi_t(a):\trho(a)\ein v$
such that whenever one of them occurrs
in $\trho(\tl)$ at $x\in\Real$ then a tile of class $a$ occurrs in
$\tl$ at $x+x_a$.
This condition allows one to locally obtain a preimage of any
representative of any $t^{-1}T'$, $T'\in[\Om]_o$.
But this furnishes also
a right inverse of $\rho$ on $\Om$ as follows:
Let $\tl$ be the representative corresponding to $\om\in\im\rho$.
Then a preimage of $\om$ under $\rho$ is given by
the pointed tiling corresponding to $\trho^{-1}(t^{-1}\tl)-x_a$ where
$t^{-1}\tl$ is rescaled in such a way that the puncture remains on $0$ and
$a$ is
determined through the unique replacement $\trho(a)$ whose puncture is on $0$.
Clearly the right inverse of $\rho$ on $\Om$ extends to $\Gs$.
And since non periodicity is forced by the existence of a locally
invertible deflation \cite{GrSh,Ke2} the corresponding \sst\ $\srho$
is locally invertible in the sense of section~\ref{11071}.

\subsection{Path spaces over graphs and their dimension groups}

This section is meant to fix the notation thereby giving an overview
on the structures that will be needed.\bs

A graph $\Sigma$ 
is a set of vertices $\Sigma^{(0)}$ and a set of edges $\Sigma^{(1)}$ with
two maps $s,r:\Sigma^{(1)}\to \Sigma^{(0)}$, the source and the range map.
Its \cm\ is the $|\Sigma^{(0)}|\times|\Sigma^{(0)}|$ matrix with
coefficients
\be
\sigma_{xy} := \mbox{number of edges which have source $y$ and range
$x$}.
\ee
A path $\xi=\xi_1\dots\xi_n$ of length $n$ over $\Sigma$ is a sequence of
$n$ edges such that $r(\xi_k)=s(\xi_{k+1})$.
We denote its lentgh by $|\xi|$. One sets
$r(\xi)=r(\xi_{n})$ and $s(\xi)=s(\xi_{1})$. Two paths $\xi,\xi'$ with
$r(\xi)=s(\xi')$
may be concatenated to yield longer path
$\xi\circ\xi'=\xi_1\dots\xi_{|\xi|}\xi'_1\dots\xi'_{|\xi'|}$.

The set of all (half) infinite paths over the graph yields a space
which carries a compact
metric topology, it is called the path space $\pf$ of the
graph.
Its topology is generated by sets
\be
U_{\xi} = \{\xi\circ\gamma|s(\gamma)=r(\xi)\}.
\ee
These sets are closed as well so that $\pf$ is a zero dimensional.
Two (infinite) paths $\gamma$, $\gamma'$ are called confinal if for some $n$,
$\forall i\geq n: \gamma_i=\gamma'_i$. Cofinality is an equivalence relation
and the subspace of $\pf\times\pf$ of all cofinal pairs may be given
a topology which is generated by
\be
\U_{\xi\xi'} = \{(\xi\circ\gamma,\xi'\circ\gamma)|s(\gamma)=r(\xi)\}
\ee
where it is required that $|\xi|=|\xi'|$ and $r(\xi)=r(\xi')$.
With this topology the groupoid defined by the
equivalence relation becomes
an $r$-discrete principal groupoid $\Gr_\Sigma$.
Its groupoid-\CA\ $\AF$ is finitely approximated, i.e.\ an $AF$-algebra.
The integer group of coinvariants of  $H(\Gr_\Sigma)$ of this groupoid is
$C(\pf,\Z)/E_\Sigma$ where $E_\Sigma$ is generated by elements of the form
$\eta(\U_{\xi\xi'}) = \Ch_{U_\xi}-\Ch_{U_{\xi'}}$.
$C(\pf,\Z)/E_\Sigma$ carries an order structure given by the notion of
positive functions, e.g.\ $[f]_\Sigma>0$ whenever it has a representative
$f>0$. With that order structure and the constant function $1$ representing
the order unit $H(\Gr_\Sigma)$
is also the dimension group or
scaled ordered $K_0$-group of the $AF$-algebra $\AF$ \cite{Eff}.
 We will not discuss the
\CA ic and $K$-theoretic details here
but only concentrate on the computation of the
dimension group which will be used lateron to determine the coinvariants of
$\Gr$.

The main point is that $H(\Gr_\Sigma)$ can be obtained by taking the algebraic
limit of the directed
system $(G_n,\sigma)$ where
$G_n=\Z^{|\Sigma^{(0)}|}$ and
$\sigma:G_n\to G_{n+1}$ is the homomorphism given by the \cm\
once the vertices have been identified with the standard base.
The algebraic limit of the above system
is a universal object which is a group $G$ together with
homomorphisms $j_n:G_n\to G$ such that $j_{n+1}\circ \sigma = j_n$ \cite{Lan}.
It is up to isomorphism determined by the property that, if there is any
other group $G'$ and $j'_n:G_n\to G'$ such that
$j'_{n+1}\circ \sigma = j'_n$ then there is a unique homomorphism
$j:G\to G'$ such that $j'_n=j\circ j_n$.
Moreover, $G$ inherits an order structure and the order unit from the
standard order structures on the $G_n$.
And $H(\Gr_\Sigma)$ coincides with $G$ as ordered group with order unit.

There are several "standard" realizations for this limit \cite{Lan,Mur}.
We will use neither of them here but instead one which is less general
but more suitable for our means, c.f.\ \cite{Hos}.
In this realization the group
is isomorphic to the quotient
\be\label{17061}
H(\Gr_\Sigma) \cong \{x\in \Real^k|\exists n\geq 0:\sigma^n(x)\in\Z^k\}/
\{x\in \Real^k|\exists n\geq 0:\sigma^n(x)=0\}
\ee
with $k=|\Sigma^{(0)}|$.
The maps $j_{n}:G_{n}=\Z^k\to H(\Gr_\Sigma)$ are given by
$j_{n}=\pi\circ \sigma^{-n}$
where $\sigma^{-1}$ is taking the preimage. If $G'$ and $j'_n:G_n\to G'$
with $j'_{n+1}\circ \sigma = j'_n$ is any other realization then
$j(\pi(x))=j'_n(\sigma^n(x))$ for $\sigma^n(x)\in\Z^k$ yields the unique
homomorphism
$j:H(\Gr_\Sigma)\to G'$ satisfying $j'_n=j\circ j_n$.
To express the order structure in this realization
we restrict for simplicity to the case where $\sigma$ is
primitive so that it has a \pfw\ $\tau$ with left-\pfv\ $\nu$.
If $\pi$ denotes the canonical projection of the above quotient,
the positive cone is
\be
H^+(\Gr_\Sigma)\cong\{\pi(x)\in H(\Gr_\Sigma)|\sum_i\nu_i x_i >0\}\cup\{0\},
\ee
and the order unit $\pi(w)\in H(\Gr_\Sigma)$ with $w_i=1$.

Since $\sigma$
is primitive, there is a unique normalized measure $\mu$ on
$\pf$ which is invariant under the groupoid $\Gr_\Sigma$,
i.e.\ satisfies
$\mu(U_\xi)=\mu(U_{\xi'})$ in case $|\xi|=|\xi'|$ and $r(\xi)=r(\xi')$
\cite{Eff}.
If the left \pfv\ of
$\sigma$ normalized to $\sum_i\nu_i=1$ then
$\mu(U_\xi)=\tau^{-|\xi|}\nu_{r(\xi)}$
and its range on $C(\pf,\Z)$ is given by
\be\label{27071}
\mu(C(\pf,\Z))
= \{\tau^{-n}\sum_i\nu_i n_i|n\geq 0,n_i\in\Z\}.
\ee
In particular $\mu$ is well defined on $H(\Gr_\Sigma)$ and the order
may be expressed as
\be           \label{11073}
H^+(\Gr_\Sigma) = \{x\in H(\Gr_\Sigma)|\mu(x)>0\}\cup\{0\} .
\ee
Elements which are neither positive nor negative are called infinitesimal.

\subsection{Path spaces determined by \sst s}

Any \sst, invertible or not,
defines an obvious graph which however does not even for all
locally invertible \sst s contain
enough information of the tiling. In analogy to the
one dimensional case we call it with \cite{For} the improper graph.
It has the \sst\ matrix as \cm.
But other graphs may also be attached to the \sst. These are related
to the improper graph and coincide with it in the border forcing case.
They allow for a coding of the tilings which yields a homeomorphism
between their path space and the hull.
This will be used to solve the
$K$-theoretical gap labelling for a certain class of \sst\ tilings.
\bs

Let for $a\in\mTei$
\be
E^n(a) = \{e\in\mTxx|R(e)=\hat{\rho}^n(a)\}.
\ee
The improper graph
$\Sigma$ is the graph which has vertices and edges resp.\
\begin{eqnarray}
\Sigma^{(0)}&=&\mTei \\
\Sigma^{(1)}&=&\bigcup_{a\in\mTei}\{(e,a)|e\in E^1(a)\}
\end{eqnarray}
and range resp.\ source maps $r,s:\Sigma^{(1)}\to\Sigma^{(0)}$ given by
\begin{eqnarray}
r(e,a)&=&a \\
s(e,a)&=&x(e).
\end{eqnarray}
Thus the edges of the graph which
have range $a$ are in bijective correspondence to the tiles in $\hat{\rho}(a)$.
$\Sigma$ has as \cm\
\be
\sigma_{a_1a_2} = |\{e\in E^1(a_1)|x(e)=a_2\}|,
\ee
i.e.\ $\sigma_{a_1a_2}$ equals the number of $a_2$'s in $\srho(a_1)$.
This matrix (or sometimes its transpose) is also called the \sst\
matrix of $\srho$.
A path of finite length like
$((e_1,a_1),\dots,(e_n,a_n))$ depends
due to the particular form of the range
map only on
$(e_1,\dots,e_n)$ and $a_n$. It shall be written shorter
$(e_1,\dots,e_n;a_n)$. \bs

The improper graph does not contain enough information of the
tiling to yield the right integer group of coinvariants.
To improve this
we have to incorporate the neighborhood of \fl s.
To a given \mixx\ $c$ let ${\F}(c)$ be the set of all possible
neighborhoods of $c$, 
i.e.\ the set of \miii es
occuring in $T$ which are composed of $c$
together with all the tiles 
the boundaries of which have
a non empty intersection of with $c$. 
The pointed tiles are those of $c$.

An integer $\mb\geq 0$ will parametrize a set of graphs
$\Laml$. (This generality is needed lateron for applications.)
Define
\be \label{30061}
\B^\mb_1:=\{(a,f)|a\in\mTei,
f\in{\F}(\srho^\mb(a)),\rho^\mb(U_a)\cap U_f\neq\emptyset\}.
\ee
The extra condition insures that only those neighborhoods of $\srho^\mb(a)$
are taken into account  which are neighborhoods of the $\mb$-fold \fl\ of $a$.
Let $\Laml$ be the graph with vertices and edges
\begin{eqnarray}
\Laml^{(0)}&=&\B^\mb_1\\
\Laml^{(1)}&=&\{(e,a,f)|(a,f)\in\B^\mb_1,e\in E^1(a)\}
\end{eqnarray}
and range and source map 
given by
\begin{eqnarray}
r(e,a,f)&=&(a,f)\\
s(e,a,f) &=& (x(e),f') \mbox{ with
$f'\ein L(\srho^\mb(e)\srho(f))$}.
\end{eqnarray}
(The $f'\in\F(\srho^\mb(a))$ for which
$f'\ein L(\srho^\mb(e)\srho(f))$ is unique.) 
The \cm\ $\laml$ of $\Laml$ has coefficients
\be
{\laml\,}_{(a_1,f_1)(a_2,f_2)} =
|\{e\in E^1(a_1)|x(e)=a_2,f_2\ein L(\srho^\mb(e)\srho(f_1))\}|.
\ee
Like for $\Sigma$ a path
$((e_1,a_1,f_1),\dots,(e_n,a_n,f_n))$ depends only on $(e_1,\dots,e_n)$
and $(a_n,f_n)$ and may be abbreviated as
$(e_1,\dots,e_n;a_n,f_n)$. Note that such a path satisfies
$R(e_k)=\hat{\rho}(x(e_{k+1}))$ and $R(e_n)=\hat{\rho}(a_n)$.
Therefore the map $\delta:\PfL^n(a,f)\to E^n(a)$ given by
\be \label{16062}
\delta(e_1,\dots,e_n;a,f) := e_1\hat{\rho}(e_2)\dots\hat{\rho}^{n-1}(e_n)
\ee
is a bijection for fixed $(a,f)$.
Furthermore
let
$\beta_\mb:\B^\mb_1\to \mTei$:
\be
\beta_\mb(a,f) = a.
\ee
Extending $\beta_\mb$ first to edges via $\beta_\mb(e,a,f):=(e,a)$ and then
to paths on obtains a map
$\beta_\mb:\PfL^n\to\pf^n$:
\be
\beta_\mb(e_1,\dots,e_n;a_n,f_n)=(e_1,\dots,e_n;a_n)
\ee
which is not only surjective but also
injective on $\PfL^n(a,f)$ for fixed $f$.
In particular $\beta_\mb$ extends to a continuous surjective map
from $\PfL$ onto $\pf$.
\bl \label{18061}
Let $T$ be a \sst\ tiling with locally invertible \sst\ $\srho$.
For each $\omega\in\Om$ there is a unique
$(e,a)\in\Sigma^{(1)}$
such that $\omega\cdot e\in\im\rho$. 
\el
{\em Proof:}
By definition of a \sst\ does for any $c\in\mTxx$ hold
the alternative:
either $c=c^{-1}$ or $\forall e_i\in E^1(x_i(c)):\:e_1\rho(c) e_2^{-1}
\neq (e_1\rho(c) e_2^{-1})^{-1}$ where $x_1(c)=x(c)$ and $x_2(c)=x(c^{-1})$.
Recall from the proof of Theorem~\ref{25071}
that $\rho(\Om)=\Om_\N$ where
$\N=\Phi(\mTxx)$.
Since for all $d\in\Phi(c)$, $\srho(c)$ and $d$ have a common greater element
the above alternative carries over
for any $d\in \N$ in the form:
either $d=d^{-1}$ or $\forall e_i\in E^1(x_i(d)):\:e_1 d e_2^{-1}
\neq (e_1 d e_2^{-1})^{-1}$.
Since $[\rho]_o(T)$ is locally isomorphic to $T$,
$\N$ has to be regular and there is a $c\in\mTxx$ such that
$\om\cdot c\in \Om_{\N}$.
And as $\N$ is approximating and generating ($T$ has to be non periodic)
$c$ can be written as $c=e d$ where $d\in\N$ and $e\in E^1(x(d))$.
We claim that $e$ is unique, for if not then
$L(c)=e d{d'}^{-1}e'^{-1}$ and since this is a unit
$d{d'}^{-1}$ must be equal to its inverse
by the above alternative. Hence $e=e'$.
Finally, if $\om\cdot ed\in\im\rho$ then also
$\om\cdot e\in\im\rho$.
The condition $a\ein \rho^{-1}(\om\cdot e)$ determines uniquely
the $a$ for which $e\in E^1(a)$.
\eb\bs

Let $\kpl:\Om\to\Laml^{(1)}$:
$\om\mapsto(e,a,f)$ where $(e,a)$ is
determined by the above lemma and $f\ein\rho^{\mb-1}(\om\cdot e)$.
Furthermore, the first component of $(e,a)$
shall be used to define
an extension of $\rho^{-1}$ to all of $\Om$ through
\be
\psi(\omega):=\rho^{-1}(\omega\cdot e).
\ee
\bl
$s(\kpl(\psi^{n}(\omega)))=r(\kpl(\psi^{n-1}(\omega))$.
\el
{\em Proof:}
Let $\kpl(\psi^{n}(\omega))=(e_2,a_2,f_2)$ and
$\kpl(\psi^{n-1}(\omega))=(e_1,a_1,f_1)$. We have to show that,
first, $x(e_2)=a_1$, and second, that the $f'\in\F(\srho^\mb(a_2))$
which satisfies
$f'\ein L(\srho^\mb(e_2)\srho(f_2))$ equals to $f_1$.
The first follows from the 
definitions: $a_1\ein\rho^{-1}(\psi^{n-1}(\om)\cdot e_1)=\psi^n(\om)$
and $L(e_2)\ein \psi^n(\om)$.
As for the second, $f_2\ein\rho^{\mb-1}(\psi^n(\om)\cdot e_2)$ implies that
$\rho(L(\srho^{\mb-1}(e_2)f_2))\ein\rho^\mb(\psi^n(\om))=
\rho^{\mb-1}(\psi^{n-1}(\om)\cdot e_1)$. Hence
$f'\ein\rho^{\mb-1}(\psi^{n-1}(\om)\cdot e_1)$ which is the relation
determining $f_1$.\eb\bs

As a consequence we may define a coding which is a map
$\Ql:\Om\to\PfL$ where the $n$th edge of $\Ql(\om)$ is
\be
\Ql(\omega)_n:= \kpl(\psi^{n-1}(\omega)).
\ee
\bt \label{16063}
Under the conditions of Lemma~\ref{18061} is $\Ql:\Om\to\PfL$ a homeomorphism.
\et
{\em Proof:}
Local invertibility implies that, for $n\geq\mb$
\be \label{23062}
\V_\mb^n(a,f,e_1,e_2) := \{[\om\cdot e_1^{-1},e_1\srho^{n-\mb}(f) e_2^{-1}]|
\om \in \rho^n(U_a)\cap\rho^{n-\mb}(U_f)\}
\ee
is compact and open for all $(a,f)\in\B^\mb_1$ and
$e_i\in E^n(a)$. (In fact, these are $\Gr$-sets.)
Let $\xi$ be a path of length $n$ with $r(\xi)=(a,f)$.
$\Ql(\om)\in U_\xi$ if and only if $L(\delta(\xi))\ein\om$, $a\in\psi^n(\om)$,
and $f\ein \rho^\mb(\psi^{n}(\om))$.
Since $\om\cdot \delta(\xi) \in
\im\rho^n$ and $\psi^n(\om)=\rho^{-n}(\om\cdot\delta(\xi))$ this means
\be \label{18062}
\Ql^{-1}(U_\xi)=\V_\mb^{|\xi|}(r(\xi),\delta(\xi),\delta(\xi)).
\ee
Hence $\Ql$ is continuous. Since
$\Ql^{-1}(U_\xi)$ is not empty, any path over $\Laml$ is an
accumulation point of $\im \Ql$. As the latter is closed $\Ql$ is surjective.
Finally, to prove injectivity,
let $r(\Ql(\om)_k)=(a_k,f_k)$ and
$\tilde{e}_n:=\delta(\Ql(\om)_1\dots \Ql(\om)_n)$.
Then $\srho^{n-\mb}(f)\ein\om\cdot\tilde{e}_n$ and
$$u_n(\om):=L(\tilde{e}_{n}\srho^{n-\mb}(f_n))\ein \om.$$
Since  $L(\tilde{e}_{n}\srho^{n}(a_n))\ein
L(\tilde{e}_{n}\srho^{n-\mb}(f_n))$
the radius of $u_n(\om)$ diverges exponentionally with $n$.
Hence $u_n(\om)$ is an approximation which approximates $\om$ and is
uniquely determined by $\Ql(\om)$.
Thus $\Ql$ is injective.
\eb\bs

Theorem~\ref{16063} implies that
$\Ql^*:C(\PfL,\Z)\to C(\Om,\Z)$: $\Ql^*(f)=f\circ \Ql^{-1}$ is an
isomorphism of groups. We may extend $\Ql$
to a map
$\Ql:\Gs\to\Gs_{\Laml}$ by restricting $\Ql\times\Ql$ to $\Gs$ and
since $T$ is not periodic we may view it as a map $\Ql:\Gr\to\Gr_{\Laml}$.
Then alike (\ref{18062})
\be
\Ql^{-1}(\U_{\xi\xi'}) =  \V^{|\xi|}_\mb(r(\xi),e_\xi,e_{\xi'})
\label{08071}
\ee
showing that this map is continuous and surjective, but it is not
injective.
Since
$\eta\circ\Ql^{-1}=\Ql^*\circ\eta$ the image  under $\Ql^*$
of the subgroup $E_{\Laml}$
is contained in $E_\Gr$. Hence there is an induced surjective
homomorphism
$[\Ql^*]:H(\Gr_{\Laml})\to H(\Gr)$.
Its kernel is $E_\Gr/\Ql^*(E_{\Laml})$.
In particular, since $\laml$ is primitive, we get a corrolary
from Theorem~\ref{16063}:
\begin{cor}
Let $T$ be a \sst\ tiling with
 primitive locally invertible \sst\ $\srho$
which reduces to a decoration of $\Z^d$, $d\leq 3$, and $\tr$ be a trace on
$A_T$. Then
\be
\tr_*(K_0(\A_\tc))
= \{\tau^{-n}\sum_i\nu_i n_i|n>0,n_i\in\Z\}
\ee
where $\tau$ is the \pfw\ and $\nu$ the left \pfv\ of
$\laml$ normalized to $\sum_i\nu_i=1$.
\end{cor}
{\em Proof:}
If $\mu$ is the measure on $\Om$ obtained by restricting the trace
then $\mu\circ\Ql^{-1}$ is a measure on $\PfL$ which is invariant under
$\Gr_{\Laml}$. Using  (\ref{06061}) one obtains
$\tr_*(K_0(\A_\tc))
= \mu(\Ql^{-1}(C(\PfL,\Z)))$ and with (\ref{27071}) the statement.\eb\bs

This solves the $K$-theoretical version of the gap labelling for
\sst\ tilings
with primitive locally invertible \sst.
\bs

{\em Remark:}
Let us only remark here that
one obtains not only the above embedding of groups by $\Ql^*$ but
in fact a unital embedding $i:\A_{\Laml}\rightarrow\A_\tc$
of the $AF$-algebra 
defined by $\Laml$ into
the algebra associated to the tiling
similar to embedding of $\AF$ in $\A_\tc$
described in \cite{Ke2}. In fact, a glance on (\ref{08071})
tells us that the characteristic functions on
$\V^{|\xi|}_\mb(r(\xi),e_\xi,e_{\xi'})$
generate $\A_{\Laml}$ as a subalgebra of $\A_\tc$.
The embedding induces an order homomorphism $i_*$ of $K$-groups.
Then $K_0(\A_{\Laml})=H(\Gr_{\Laml})$
as ordered group with order unit, and, under the hypothesis
(\ref{08075}),
 $i_*:K_0(\A_{\Laml})\to H(\Gr)$ coincides with $[\Ql^*]$.
However, in case $i_*$ is not surjective, i.e.\ the dimension of the tiling
is bigger than one, our analysis will not determine the order structure on
$K_0(\A_\tc)$ but only a subcone of the positive cone $K^+_0(\A_\tc)$.

\subsection{The integer group of coinvariants for \sst\ tilings}

Recall that the integer group of coinvariants is given by
$H(\Gr)=H(\Gr_{\Laml})/\ker [\Ql^*]$.
To compute $\ker [\Ql^*]=E_\Gr/\Ql^*(E_{\Laml})$ we need to control \miii es
consisting of two tiles. In analogy to (\ref{30061},\ref{23062})
define\footnote{In principle the $\mb$ below could be chosen different from
the one above, but we shall not make use of this generality.}
for $n\geq\mb$
\be
\B^\mb_{2\neq}:=
\{(A,F)|A\in\mTzw,F\in\F(\srho^\mb(A)),\rho^\mb(U_A)\cap U_F\neq\emptyset\}.
\ee
\be \label{28061}
\V_\mb^n(A,F,e_1,e_2) := \{[\om\cdot e_1^{-1},e_1\srho^{n-\mb}(F) e_2^{-1}]|
[\om,\srho^{n-\mb}(F)] \in \rho^n(U_A)\cap\rho^{n-\mb}(U_F)\}
\ee
where $e_i\in E^{n}(x_i(A))$, $x_1(c)=x(c)$ and $x_2(c)=x(c^{-1})$.
If $e_i$ are units we write $\V_\mb^n(A,F)$ and resp.\
$\V_\mb^n(a,f)$ for the above.
It should be clear that
for fixed $e_1,e_2$ the sets $\V_\mb^n(x,y,e_1,e_2)$ are for different
$(x,y)\in\B^\mb_1\cup\B^\mb_{2\neq}$ pairwise disjoint.
$[\,\cdot\,]_{\Laml}$ denotes equivalence classes
with respect to $\Ql^*(E_{\Laml})$. \bs
\bl \label{07075}
Let $\tc$ be a \sst\ tiling with locally invertible \sst\ and $\Laml$ and $\Ql$
as above. Then
$E_\Gr/\Ql^*(E_{\Laml})$ is generated by elements of the form
$\left[\eta(\V_\mb^n(A,F))\right]_{\Laml}$
for $(A,F)\in\B^\mb_{2\neq}$.
\el
{\em Proof:} Since $\mTzw$ generates $\mTxx\backslash\mTei$
a generating set for $E_\Gr$ is provided by the set of
elements of the form
$\eta(\Ch_{c})$ where $c=uA$, $u\in\mTx$, $u\cp A\in\mTzw$.
For such a $c$ let
\begin{eqnarray*}
I_1^{n}(c)&:=&\{(a,f,e_1,e_2)|(a,f)\in\B^\mb_1,e_i\in E^{n}(a)|
\exists c':c, e_1 \srho^{n-\mb}(f) e_2^{-1}\ein c'\}\\
I_{2\neq}^{n}(c)&:=&\{(A,F,e_1,e_2)|(A,F)\in\B^\mb_{2\neq},e_i\in
E^{n}(x_i(A))|
\exists c':c, e_1 \srho^{n-\mb}(F) e_2^{-1}\ein c'\}
\end{eqnarray*}
Consider $\om\in U_{L(c)}$ and let $n\geq \mb$.
By Lemma~\ref{18061} there is a unique
$(a,f,e_1)$ such that $\om\in\V_\mb^n(a,f,e_1,e_1)$.
It follows that $e_1^{-1}\cp c$. Now suppose that
$\exists e_2\in E^n(a)\exists c':e_1e_2^{-1}, c\ein c'$. Then this $e_2$
is uniquely determined and
$[\om,c]\in \V_\mb^n(a,f,e_1,e_2)$.
In particular  $\V_\mb^n(a,f,e_1,e_2)$  are for $(a,f,e_1,e_2)\in I_1^{n}(c)$
pairwise disjoint.
If the above assumption is not satisfied then, because of the
form $c=uA$, there must be a $(A,F)$ with $a=L(A)$ and $f\ein L(F)$
such that
$\exists e_2\in E^n(x_2(A))\exists c':e_1\srho^{n-\mb}(F)e_2^{-1}, c\ein c'$.
Again $e_2$ is uniquely determined
and $[\om,c]\in \V_\mb^n(A,F,e_1,e_2)$. This shows that
$\V_\mb^n(A,F,e_1,e_2)$ are for $(A,F,e_1,e_2)\in I_{2\neq}^{n}(c)$
pairwise disjoint and hence
\be \label{30064}
\U_c \subset
\dot{\bigcup_{I^n_1(c)}}\V_\mb^n(a,f,e_1,e_2)\:\dot{\cup}\:
\dot{\bigcup_{I_{2\neq}^n(c)}}\V_\mb^n(A,F,e_1,e_2).
\ee
Since for all $(x,y)\in\B^\mb_1\cup\B^\mb_{2\neq}$ and $e_i\in E^{n}(x_i(x))$,
$\rad(L(e_1\srho^{n-\mb}(y)))$ 
diverges
there is an
$n$ such that
for all $y$ and all $e_i$ either $c\ein e_1\srho^{n-\mb}(y)e_2^{-1}$
or $\not\exists c':c, e_1 \srho^{n-\mb}(y) e_2^{-1}\ein c'$.
In other words,
either $\U_c\cap\V^n_\mb(x,y,e_1,e_2)=\V^n_\mb(x,y,e_1,e_2)$
or that intersection is empty. Thus the inclusion in (\ref{30064})
is an equality for large enough $n$.

By (\ref{08071}),
$\V_\mb^n(a,f,e_1,e_2)\in \Ql^{-1}(\asg(\Gr_{\Laml}))$ so that
$[\eta(\V_\mb^n(a,f,e_1,e_2))]_{\Laml}=0$. Thus
the first part of the union (\ref{30064}) will not contribute.
As for the second, observe that
\be\label{08072}
L(\V_\mb^n(A,F,e_1,e_2))=\{\om\cdot e_1^{-1}|
\om\in\rho^n(U_{L(A)})\cap\rho^{n-\mb}(U_{L(F)})\}
\ee
and
$\V_\mb^n(a,f,e,e)=\{\om\cdot e^{-1}|
\om\in\rho^n(U_{a})\cap\rho^{n-\mb}(U_{f})\}$. As above, if $l$ is big enough
one gets the alternative
\be\label{30063}
 L(\V_\mb^n(A,F,e_1,e_2))\cap \V^{n+l}_1(a,f,e,e)=\left\{
\begin{array}{l}
L(\V_\mb^n(A,F,e_1,e_2)) \\
\emptyset \end{array}\right..
\ee
It follows that (\ref{08072}) (which does not depend on
$e_2$)
can be written as a disjoint union
$$L(\V_\mb^n(A,F,e_1,e_2))=\dot{\bigcup_{(a,f,e)\in J^l(A,F,e_1)}}
\V^{n+l}_1(a,f,e,e)$$
for an appropriate $J^l(A,F,e_1)$, and moreover,
$J^l(A,F,ce_1)$ with $R(c)=L(e_1)$ equals to
$\{(a,f,ce)|(a,f,e)\in J^l(A,F,e_1)\}$. Thus
$[L(\V_\mb^n(A,F,e_1,e_2))]_{\Laml}=[L(\V_\mb^n(A,F))]_{\Laml}$ and we
end up with
\be
[\eta(\U_c)]_{\Laml} = \sum_{(A,F,e_1,e_2)\in
I_{2\neq}^n(c)}[\eta(V^n_\mb(A,F))]_{\Laml}.
\ee
Since $\eta(V^n_\mb(A,F))\in E_\Gr$
the lemma is proven.\eb

\bl
Let $(a,f)\in\B^\mb_1$, $(A,F)\in\B^\mb_{2\neq}$, $e\in E^{l}(a)$,
$n\geq \mb$,
and $l\geq \mb$ large enough  for the alternative
(\ref{30063}) to hold. Then $L(\V_\mb^n(A,F))
\cap \V_\mb^{n+l}(a,f,\srho^n(e),\srho^n(e))\neq\emptyset$ whenever
$L(\srho^\mb(e^{-1})F)\ein\srho^l(f)$ and $L(e^{-1}A)\ein\srho^{l-\mb}(f)$.
\el
{\em Proof:}
$$ L(\V_\mb^n(A,F))
\cap \V_\mb^{n+l}(a,f,\srho^n(e),\srho^n(e))\neq\emptyset
$$
is for $n\geq \mb$ equivalent to
$$
\{\om\cdot \srho^\mb(e) |\om\in\rho^\mb(U_{L(A)})\cap U_{L(F)}\}
\cap \rho^{l+\mb}(U_a)\cap \rho^l(U_f)
\neq\emptyset.
$$
That the latter implies
$L(\srho^\mb(e^{-1})F)\ein\srho^l(f)$ and $L(e^{-1}A)\ein\srho^{l-\mb}(f)$
is clear provided $l$ is large enough.
Now let $(a,f)\in\B^\mb_1$, $(A,F)\in\B^\mb_{2\neq}$, and
$\om\in  \rho^{l+\mb}(U_a)\cap \rho^l(U_f)$. Then
$L(\srho^\mb(e^{-1})F)\ein \srho^l(f)$ implies
$L(F)\ein \om\cdot\srho^\mb(e^{-1})\in \im\rho^\mb$, and
$L(e^{-1}A)\ein\srho^{l-\mb}(f)$ implies
$L(A)\ein \rho^{-\mb}(\om\cdot\srho^\mb(e^{-1}))$. \eb\bs

Define the $|\B^\mb_1|\times |\B^\mb_{2\neq}|$ matrices
with coefficients $\Kl_{(a,f)(A,F)}$ resp.\ $\KKl_{(a,f)(A,F)}$
through
\be \label{07021}
\Kl_{(a,f)(A,F)}:= |\{e\in E^{l}(a)|
L(\srho^\mb(e^{-1})F)\ein\srho^l(f), L(e^{-1}A)\ein\srho^{l-\mb}(f)
\}|.
\ee
with $l\geq \mb$ and
\be
\KKl_{(a,f)(A,F)}:=\Kl_{(a,f)(A,F)}-\Kl_{(a,f)(A^{-1},F^{-1})}.
\ee
\bt
Let $\tc$ be a \sst\ tiling with locally invertible \sst\ and
$\laml$ and $\KKl$ as above.
If $l$ is large enough so that the alternative (\ref{30063}) holds then,
with $k=|\B^\mb_1|$,
\be
H(\Gr)\cong \{x\in \Real^k|\exists n\geq 0:\laml^n(x)\in\Z^k\}/
\{x\in \Real^k|\exists n\geq 0:\laml^n(x)\in\im\KKl\}.
\ee
\et
{\em Proof:}
By the last lemma
$$
[\Ch_{L(\V_\mb^n(A,F))}]_{\Laml} = \sum_{(a,f)\in\B^\mb_1}
 [\Ch_{\V_\mb^{n+l}(a,f)}]_{\Laml} \Kl_{(a,f)(A,F)}
$$
provided alternative (\ref{30063}) holds, and hence
\be \label{07071}
[\eta({\V_\mb^n(A,F)})]_{\Laml} = \sum_{(a,f)\in\B^\mb_1}
 [\Ch_{\V_\mb^{n+l}(a,f)}]_{\Laml} \KKl_{(a,f)(A,F)}.
\ee
As we already saw,
$[\Ch_{\V_\mb^{n}(a,f)}]_{\Laml}=[\Ql^*(\Ch_\xi)]_{\Laml}$ with $|\xi|=n$
and $r(\xi)=(a,f)$ so that these elements generate
$H(\Gr_{\Laml})$.
Moreover
\be \label{07022}
[\Ch_{\V_\mb^{n}(a,f)}]_{\Laml} = \sum_{(a',f')}
[\Ch_{\V_\mb^{n+1}(a,f)}]_{\Laml} {\laml\,}_{(a',f')(a,f)}
\ee
so that
\be \label{30062}
\KK^{(l+n,\mb)}=\laml^n \KKl.
\ee
Recall (\ref{17061}) that
$$H(\Gr_{\Laml}) = \{x\in \Real^k|\exists n\geq 0:\laml^n(x)\in\Z^k\}/
\{x\in \Real^k|\exists n\geq 0:\laml^n(x)=0\}.$$
The image of $E_\Gr/\Ql^*(E_{\Laml})$ in $H(\Gr_{\Laml})$
is by Lemma~\ref{07075} and (\ref{07071},\ref{30062})
generated by the images of
$j_n\circ\KKl=\pi\circ\laml^{-n}\circ\KKl$.
Now
$\pi(x)\in\im j_n\circ\KKl$ whenever
$\laml^n(x)\in\im\KKl$ for some
representative $x$ of $\pi(x)$. \eb\bs

Suppose that $\Om$ carries an $\Gr$-invariant (normalized) measure and that the
\sst\ is primitive. Then this measure is unique (and the tiling minimal
\cite{Ke2}) and the order structure of
$H(\Gr)$ can be expressed  as
$[x]>0$ for $[x]\in H(\Gr)$ if and only if $\sum_i\nu_ix_i>0$
where $\nu$ is the left-\pfv\ of $\laml$. In particular the latter
inequality is independent of the chosen representative
(the elements of $\ker[\Ql^*]$ are infinitesimal).

\subsubsection{Simplifications}

The determination of $\KKl$ can be quite cumbersome.
But simplifications occurr under certain circumstances. \ms

1) In case that for all $(a,f)\in\B^\mb_1$, $a$ is uniquely determined by $f$
(we write it as $a(f)$) $L(\srho^\mb(e^{-1})F)\ein \srho^l(f)$ implies
$L(e^{-1}A(F))\ein\srho^{l-\mb}(f)$.
Thus $\Kl$ simplifies to
\be
\Kl_{fF} :=
 \Kl_{(a,f)(A,F)} =
|\{e\in E^{l}(a(f))|L(\srho^\mb(e^{-1})F)\ein \srho^l(f)\}|.
\ee
This is for instance the case if $\mb=0$.

2) In some sense the other extreme is that the $f$ is determined by $a$.
This is case if the \sst\ forces its border.
It has been partly analyzed in \cite{Ke2}. In the present notation it
is expressed as follows:
\bd
A locally invertible \sst\ forces its border if there is a $\mb$
such that $\beta_\mb:\B^\mb_1\to\mTei$ is a bijection.
\ed
Not only the graph simplifies enormously, in that it coincides with the
improper graph, but also we may take $l=\mb$.
Moreover $L(e^{-1}A)\ein\srho^{l-\mb}(f(a))$
implies $L(\srho^\mb(e^{-1})F(A))\ein\srho^l(f(a))$ where we wrote
$f(a)$ for the $f$ determined by $a$. Hence for \sst s which force
their border
\be
K^{(\mb,\mb)}_{aA}:=  K^{(\mb,\mb)}_{(a,f)(A,F)} =
|\{e\in E^{\mb}(a)|L(e^{-1}A)\ein f(a)\}|.
\ee
Note that a \sst\ can force its border only for $\mb\geq 1$, because otherwise
any pointed tiling would be determined by its pointed tile and hence
periodic which contradicts local invertibility. \ms

As for the determination of $H(\Gr)$ we have:\ms

3) If $\laml$ is invertible over $\Z$ then
$H(\Gr)=\Z^k/\langle\bigcup_n
\laml^{-n}\im\KKl\rangle$.

\subsection{Examples}

The above machinerie has been designed to tackle higher dimensional
tilings. We therefore will present the computation of the coinvariants
associated to the Penrose tilings.
But it is also applicable to one dimensional \sst s
in the sense of \cite{Q}.
To allow the reader a comparison with technics used elsewhere to obtain
the integer group of coinvariants, c.f.\ \cite{For,Hos}, we present the
Thue Morse \sst\ as an example.

\subsubsection{The Thue Morse \sst}
The Thue Morse \sst\ $\varrho$ is defined on the two letter alphabet $\{a,b\}$
by
\begin{eqnarray}
a &\mapsto& ab\\
b &\mapsto& ba
\end{eqnarray}
and extended to words as
$\varrho(a_1\dots a_k)= \varrho(a_1)\dots \varrho(a_k)$.
It may be viewed (like any other one dimensional substitution of the kind
in \cite{Q}) as a \sst\ in the algebraic sense of the tiling which is
a fixed point under $\varrho$:
Consider the sequence
over $\Z^{\geq 0}$ with values in $\{a,b\}$ given by
$\varrho^\infty(a)$ and
complete it to a sequence over $\Z$ by reflection
(i.e.\ by $\varrho^{2\infty}(a)$ to the left).
This represents
a one dimensional pointed tiling the pointed tile (letter) being
the one on $0$, i.e.\
the first one of (the right) $\varrho^\infty(a)$.
The (finite) words appearing in $\varrho^\infty(a)$
with two letters chosen are \mixx es.\footnote{
If we do not have a geometric interpretation of the tilings as
sequences of decorated intervals we have to restrict to \miii es which
are connected.}
We indicate the first by
a grave and the second by an acute, and if both coincide by a check.
(An example of a multiplication is
$(\lp{a}\fp{b})(b\fp{a}b\lp{b})=(ba\fp{b}\lp{b}))$.
A \sst\ as a homomorphism of the \AG\ of \mixx es is then given by
$\srho(w_1\fp{a}_iw_2\lp{a}_jw_3)=\varrho(w_1)\fp{a}_{i_1}a_{i_2}
\varrho(w_2)\lp{a}_{j_1}a_{j_2}\varrho(w_3)$ where we have used the writing
$\varrho(a_i)=a_{i_1}a_{i_2}$.

The \sst\ is locally invertible but does not force its border and
there is no advantage in using large $\mb$. We therefore take $\mb=0$.
The \cm\ of $\Sigma_0$ is
$${\lambda_0\,}_{f_1f_2} =\mbox{number of $e\in E^1(a(f_1))$ with
$f_2\ein L(e\crho(f_1))$}. $$
We have
$$\B^0_1=\{a\bp{a}b,b\bp{a}a,b\bp{a}b,b\bp{b}a,a\bp{b}b,a\bp{b}a\}$$
$$\srho(\B^0_1)=\{ab\bp{a}bba,ba\bp{a}bab,ba\bp{a}bba,
ba\bp{b}aab,ab\bp{b}aba,ab\bp{b}aab\}$$
$$E^1(a)=\{\bp{a}b,\lp{a}\fp{b}\}\quad E^1(b)=\{\bp{b}a,\lp{b}\fp{a}\}$$
Taking the elements of $\B^0_1$ and $\srho(\B^0_1)$ in the above order one
obtains
$$\lambda_0 =
\left(\begin{array}{cccccc}
0 & 0 &1 &0 &1 &0 \\
1 & 0 &0 &0 &0 &1 \\
1 & 0 &0 &0 &1 &0 \\
0 & 1 &0 &0 &0 &1 \\
0 & 0 &1 &1 &0 &0 \\
0 & 1 &0 &1 &0 &0
\end{array}\right).
$$
The \pfw\ is $2$ and the normalized left-\pfv\ is
$\nu=\frac{1}{6}(1,1,1,1,1,1)$. Hence
\be
\mu(H(\Gr)) = \frac{1}{3}\Z[\frac{1}{2}].
\ee
To obtain the full group we may split up
$\B^0_{2\neq}=\B^0_{2<}\cup (\B^0_{2<})^{-1}$ with
$$\B^0_{2<}=\{a\fp{a}\lp{b}a,a\fp{a}\lp{b}b,b\fp{a}\lp{b}a,b\fp{a}\lp{b}b,
b\fp{a}\lp{a}b,b\fp{b}\lp{a}b,b\fp{b}\lp{a}a,a\fp{b}\lp{a}b,a\fp{b}\lp{a}a,
a\fp{b}\lp{b}a\}.$$
Thus (\ref{30063}) holds for $l=1$, i.e.\ we have to determine
$$K^{(1,0)}_{fF} =\mbox{number of $e\in E^1(a(f))$ with
$L(F)\ein L(e\srho(f))$}. $$
Let $ \KK^{(1,0)}_{<}$ be the restriction of $\KK^{(1,0)}$ to indices
$fF$ with $F\in\B^0_{2<}$. Then, again with respect to the above order,
$$ \KK^{(1,0)}_{<}=
\left(\begin{array}{cccccccccc}
0 & 0 &0 &0 &0 &0 &0 &-1 &0 &1\\
0 & 0 &0 &0 &-1 &0 &0 &1 &0 &0\\
0 & 0 &0 &0 &-1 &0 &0 &0 &0 &1\\
0 & 0 &-1 &0 &1 &0 &0 &0 &0 &0\\
0 & 0 &1 &0 &0 &0 &0 &0 &0 &-1\\
0 & 0 &0 &0 &1 &0 &0 &0 &0 &-1
\end{array}\right).
$$
And, since $\KK^{(1,0)}_{fF^{-1}} = - \KK^{(1,0)}_{fF}$,
$\im\KK^{(1,0)}=\im\KK^{(1,0)}_{<}\cong\Z^3$.
$\lambda_0\KK^{(1,0)}_{<}$ is up to a permutation of the columns 
$\KK^{(1,0)}_{<}$, i.e.\ $\lambda_0$ preserves $\im\KK^{(1,0)}$.
In fact, $\lambda_0$ is diagonalizable, it has eigenvalues $2,1,0,-1$,
$-1$ occurring with multiplicity $3$, and $\im\KK^{(1,0)}$ is spanned by
the (right) eigenvector to eigenvalue $1$ together with a two dimensional
subspace of the eigenspace to eigenvalue $-1$.
A system of (right) eigenvectors is given by
$$
\left(\begin{array}{c} 1\\1\\1\\1\\1\\1 \end{array}\right)\quad
\left(\begin{array}{c} -1\\0\\-1\\1\\0\\1 \end{array}\right)\quad
\left(\begin{array}{c} -1\\-1\\-1\\1\\1\\1 \end{array}\right)\quad
\left(\begin{array}{c} 0\\-1\\-1\\0\\1\\1 \end{array}\right)\quad
\left(\begin{array}{c} -1\\1\\0\\-1\\1\\0 \end{array}\right)\quad
\left(\begin{array}{c} 1\\-1\\-1\\1\\0\\0 \end{array}\right).
$$
It follows that
\be
H(\Gr)\cong\Z[\frac{1}{2}]\oplus\Z.
\ee
The elements of the second summand are infinitesimal
since the pairing between the left \pfv\ $\nu$ and all vectors from the
(right) eigenspace to eigenvalue $-1$ is zero. Thus
the positive cone is $\Z[\frac{1}{2}]^+$.

\subsubsection{Penrose tilings} \label{03081}

There are several well known
variants of tilings which are called Penrose tilings \cite{Pen,GrSh}
and which are a priori to be distinguished as they
lead to non-isomorphic groupoids. But they may be transformed
into each other by purely local manipulations
which implies that one can find maps satisfying the conditions
of Lemma~\ref{27061} and leading to isomorphisms between reductions of the
corresponding groupoids. Since any such tiling is minimal
all reductions lead to the same ordered integer group of coinvariants
differing possibly in the order unit.
The version which is most suitable for our purposes is the one which has
triangles as \ti s, cf.\ Figure~7.
The triangles are decorated (with a little circle) to break the mirror
symmetry. There are $40$ \miii es of them.
The orientational symmetry of a tiling (or its class)
is the largest subgroup of $O(d)$,
acting on a pointed tiling being identified with a representative in
$\Real^d$ in the obvious way, which leaves the hull invariant.
The orientational symmetry of a Penrose tiling by triangles
possesses $20$ elements. It is
generated by a rotation around $\frac{\pi}{5}$ together with a
mirror reflection at a boundary line of a triangle \cite{Ke2}.
We denote by $\mTxx$ resp.\ $\Gr$ the \AG\ resp.\
groupoid associated to these tilings.
The well known deflation of these tilings
$\trho$ with $t=\frac{1+\sqrt{5}}{2}$ is displayed below.
Since it is covariant with respect to the orientational symmetry
it suffices to give it for one orientation only.

\epsffile[0 0 430 170]{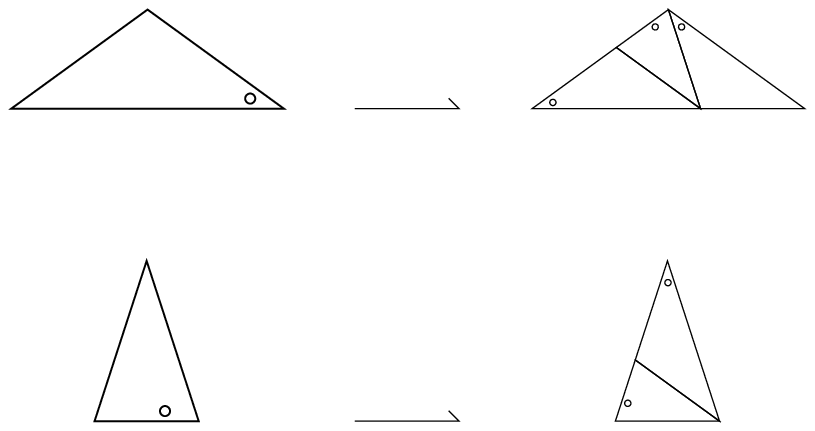}

A replacement $\trho(a)$ taking exactly the same space as the tile $a$ the
relative position measured with respect to the centers of gravity is $0$.
Choosing pointed tiles for $\trho(a)$ we obtain a \sst\ $\srho$.
Those tilings which have an exact five-fold
symmetry are invariant under $\srho^4$.
The \sst\ is primitive, locally invertible,
and forces its border with $\mb=4$ \cite{Ke2}.
Hence $\Lambda_4=\Sigma$ and $\lambda_4=\sigma$, the \sst\ matrix of
$\srho$.
Since $\sigma$ is invertible over the integers $H(\Gr_\Sigma)\cong\Z^{40}$.
To simplify the computation of $E_\Gr/Q_4^*(E_\Sigma)$ we make
use of the symmetry properties of the tiling and in particular of the fact
that the boundaries of \fl s $\srho^4(a)$ are local mirror axes so that
the pattern classes of those $A\in\mTzw$ which cross the boundaries of
$4$-fold \fl s are always mirror symmetric \cite{Ke2}, c.f.\ below where
the boundaries are indicated through fatter lines.

\epsffile[0 0 430 190]{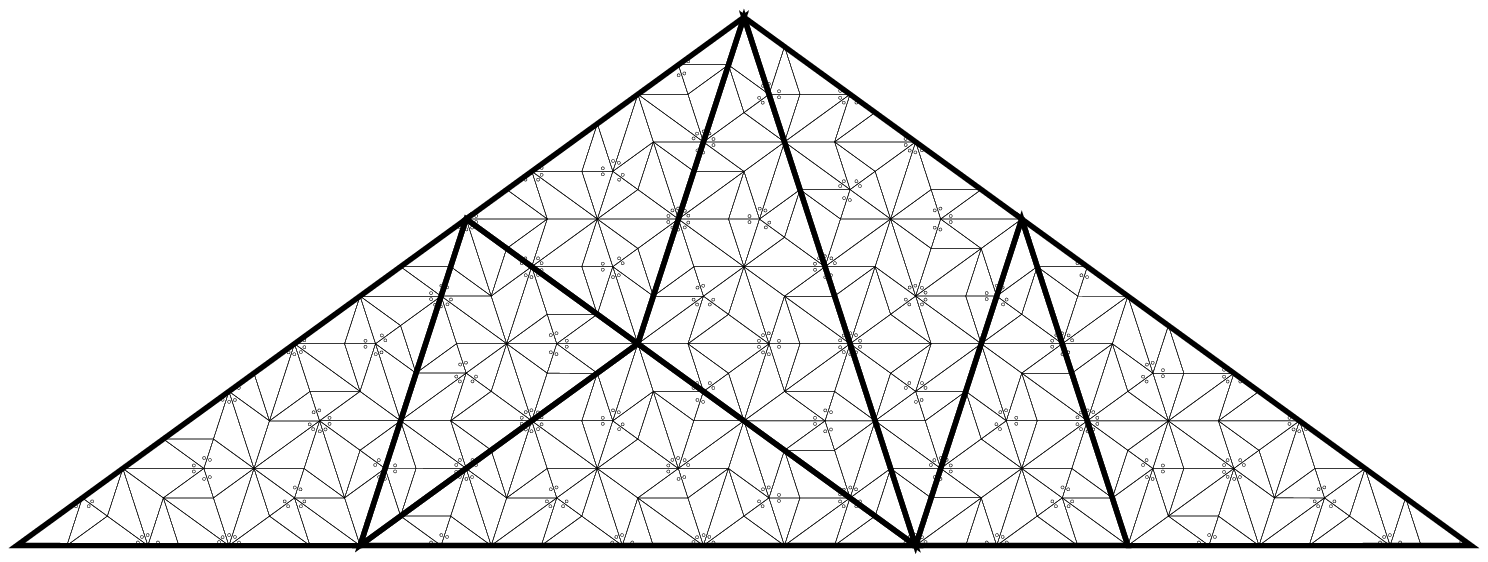}

Let $\eg$ denote the direction of a boundary line of a tile resp.\ a \fl.
There are
$10$ different ones and we order them  anti-clockwise identifying them with
$\{0,\cdots,9\}$. Saying that a \fl\ has boundary $\eg$ if it has a boundary
with that direction we define the $40\times 40$ matrices with entries
\be
N^\eg_{ab} := \mbox{number of $a$'s at boundary $\eg$ of $\srho^4(b)$} .
\ee
In particular $N^\eg_{ab}=0$ in case $a$ or $\rho(b)$ do not have
boundary $\eg$. Let $\eg(a)$ be the mirror
image of $a$ with respect to the mirror axis $\eg$ and define
\be
{\cal D}^\eg_{ab} = N^\eg_{ab}-N^\eg_{a \eg(b)}.
\ee
Then
\be
\KK^{(4,4)}_{a A} = \left\{
\begin{array}{cl}
{\cal D}^\eg_{ab} & \mbox{if } |A|=b \eg(b)\mbox{ and } x(A)=b \\
-{\cal D}^\eg_{ab} & \mbox{if } |A|=b \eg(b)\mbox{ and }x(A)=\eg(b)\\
0 & \mbox{else}
\end{array}
\right.
\ee
$|A|=b\eg(b)$ indicating that the pattern class of $A$
is composed of $b$ and $\eg(b)$
in such a way that the common boundary is the symmetry axis.
Hence
$\im \KK^{(4,4)}=\erz{\im{\cal D}^\eg,\eg=0,\dots,9}$.
${\cal D}^\eg$ is related to ${\cal D}^{0}$ by symmetry, i.e.\
${\cal D}^\eg=R^{-\alpha}{\cal D}^{0} R^\alpha$,
$R$ being the matrix which acts as a rotation around $\frac{\pi}{5}$.

To be very explicit let us use a basis
$\{e_{10 k+\eg}\}_{0\leq k \leq 3, 0\leq\eg\leq 9}$ of
$H(\Gr_\Sigma)$ with
$e_{10 k+\eg}=[\Ch_{U_{a_{10 k+\eg}}}]_\Sigma$
where $a_{10 k+\eg}$ corresponds to the \miii\ of the
 triangle in Figure~7.k 
rotated around an angle of $\frac{\alpha\pi}{5}$.

\epsffile[0 0 430 100]{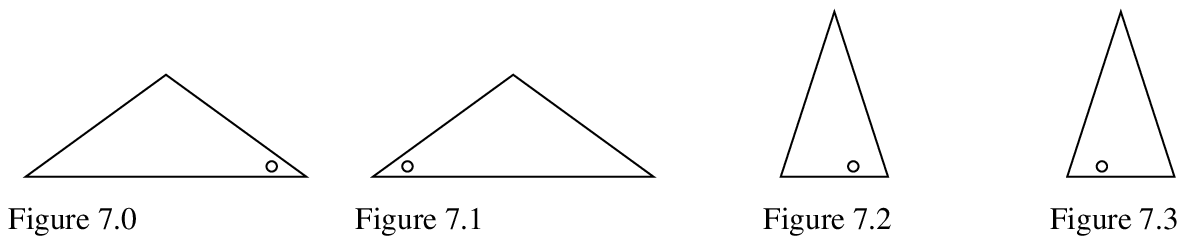}

In terms of the rotation matrix $\omega$, which has
entries $\omega_{\alpha\beta}=\delta_{\alpha-\beta,1\;mod\;10}$,
$R$ and the \sst\ matrix $\sigma$ are given by
$$
R  =  \left(
\begin{array}{cccc}
\omega & 0 & 0 & 0 \\
0 & \omega & 0 & 0 \\
0 & 0 & \omega & 0 \\
0 & 0 & 0 & \omega
\end{array}
\right), \quad
\sigma  =  \left(
\begin{array}{cccc}
\omega^4 & \omega^0 & 0 & \omega^6 \\
\omega^0 & \omega^6 & \omega^4 & 0 \\
\omega^3 & 0 & \omega^7 & 0 \\
0 & \omega^7 & 0 & \omega^3
\end{array}
\right).
$$
The matrix $N^{0}$ may be read of from Figure~8 
(and completed by symmetry); it is given below (\ref{19011}).
Moreover ${\cal D}^0=N^0-N^0S$ where $S$ implements the reflection at
$\eg=0$, explicitly, with
$s_{\alpha\beta}=\delta_{\alpha+\beta,5\;mod\;10}$ (counting rows and
columns form $0$ to $9$)
$$
S  =  \left(
\begin{array}{cccc}
0 & s & 0 & 0 \\
s & 0 & 0 & 0 \\
0 & 0 & 0 & s \\
0 & 0 & s & 0
\end{array}
\right).
$$
It turns out that $\im\KK^{(4,4)}$ is generated by
the orbit under $R$ of (the transpose of) the four vectors
$$\begin{array}{c}
v_1=(0,\!0,\!0,\!0,\!0,\!0,\!0,\!0,\!0,\!0,\!0,\!0,\!0,\!0,\!0,\!0,\!0,\!0,\!0,
\!0,\!0,\!0,\!0,\!0,\!
   0,\!0,\!1,\!0,\!0,\!0,\!0,\!-1,\!0,\!0,\!0,\!0,
\!0,\!0,\!0,\!0)\\
v_2=(0,\!0,\!0,\!0,\!0,\!0,\!1,\!0,\!0,\!0,\!0,\!-1,\!0,\!0,\!0,
\!0,\!0,\!0,\!0,\!0,\!0,\!0,\!0,\!0,\!
   0,\!0,\!0,\!0,\!0,\!0,\!0,\!0,\!0,\!0,\!0,\!0,\!0,\!0,\!0,\!0)\\
w_1=(0,\!0,\!0,\!0,\!0,\!0,\!0,\!1,\!0,\!0, -1,\!0,\!0,\!0,\!0,
\!0,\!0,\!0,\!0,\!0,\!0,\!0,\!0,\!0,\!
   0,\!0,\!0,\!0,\!1,\!0,\!0,\!0,\!0,\!0,\!0,\!0,\!0,\!0,\!0, -1)\\
w_2=(0,\!0,\!0,\!0,\!0,\!1,\!0,\!0,\!0,\!0,\!0,\!0, -1,\!0,
\!0,\!0,\!0,\!0,\!0,\!0,\!0,\!0,\!0,\!0,\!
   0,\!0,\!0,\!0,\!0, -1,\!0,\!0,\!0,\!0,\!0,\!0,\!0,\!0,\!1,\!0)
\end{array}
$$
but $8$ of the $40$ vectors thus obtained are linearly dependent of the others.
Moreover $\im\KK^{(4,4)}$ is invariant
under $\sigma$. Dividing it out yields no torsion so that we obtain
\be\label{31071}
H(\Gr)\cong\Z^8.
\ee
To obtain the order structure we look for the smallest subgroup $I$ of
$\Z^{40}$
which is invariant under $\sigma$ and spans a real vector space containing
the right \pfv\ of $\sigma$. Then the positive cone
of $\Z^{40}$ is given by those $x\in I$ which pair with the left-\pfv\
$\nu$ of $\sigma$ to $\nu x >0$.
All other elements must pair with $\nu$ to zero and are thus infinitesimal.
Clearly $I$ is spanned by (the transpose of)
$$ x_1 = (\overbrace{1,\dots,1}^{20},\overbrace{0,\dots,0}^{20}),
\quad x_2 = (\overbrace{0,\dots,0}^{20},\overbrace{1,\dots,1}^{20}) $$
and $\nu$ is given by $\nu = \frac{2-t}{20}(tx_1+x_2)^T$ (in its normalized
form).
Thus $I\cap \im\KK^{(4,4)}=\emptyset$ and
\be\label{03083}
H(\Gr)\cong I \oplus \Z^6
\ee
the elements of the second summand being infinitesimal.
The range of the unique state on $H(\Gr)$, which
of course coincides with that on $H(\Gr_\Sigma)$, is
\be \label{03082}
\mu(H(\Gr))=\frac{1}{20}(\Z + t\Z).
\ee
This concludes the computation of the integer group of coinvariants and
the range of its state for the Penrose tilings.
But it may be instructive to
 look at result (\ref{31071}) from a different point of view.
Let $H_1$ be the sublattice of $\Z^{40}$
which is generated by $\{R^\eg v_i\}_{i=1,2;\,\eg=0,1\cdots,9}$.
$H(\Gr_\Sigma)/H_1=\Z^{20}$ has a basis with natural geometrical
interpretation.
It is formed by the classes of
characteristic functions on the \miii es of rhombi which
are always formed by either
two smaller triangles or two larger triangles.
This indicates that
one should look at the Penrose tilings by rhombi.
Such a tiling, which we denote by $T_{Rh}$, has only $20$ \miii es
of \ti s.
That the groupoid 
associated to
the Penrose tilings by rhombi is isomorphic to a reduction of $\Gr$ may be
seen as follows:
Let $\C\subset\mTei$ consist of the classes of triangles
that are obtained from those of Figure~7.1 and 7.3 by a rotation.
Let $\Rh\subset\mTxx$ be the \AG\ given by elements of the form
$aca'$ where $a,a'\in\C$ and $c\in\mTxx$ is such that
all its triangles pair to rhombi. Clearly $\Rh$ is approximating generating
and regular.
Now deleting the diagonal
which coincides with the base of the two touching triangles
yields an isomorphism
of \AG s $\Rh\to\mtxx{T_\Rh}$ which satifies the growth condition with
$t=1$. Hence
the groupoid associated to the rhombus version is isomorphic to $\Gr_\Rh$.

The largest subgroup of $O(2)$ leaving $\Om_\Rh$ invariant
consists only of the  rotations which are multiples of $\frac{\pi}{5}$
since the mirror image of a $\om\in\Om_\Rh$ lies in $\Om\backslash\Om_\Rh$.
However, to our knowledge there is no \sst\ for Penrose
tilings by rhombi which is covariant even under this
reduced orientational symmetry.
But there are non covariant ones,
namely $10$ of them, the deflation
corresponding to $\srho_0$ being given in Figure~9 
and $\srho_\eg$
being obtained from $\srho_0$ just by rotation of the whole figure
around $\frac{\eg\pi}{5}$, $\eg=0,\cdots,9$.
(The relative positions are indicated by a cross.)
That all these
\sst s are primitive locally invertible and force their \saum\ carries over
from $\srho$.
It turns out that
\be
\ker\sigma_\eg = \erz{\{[R^\eg w_i]_{H_1}\}_{i=1,2}}
\ee
where $\sigma_\eg$ is the \sst\ matrix of $\srho_\eg$ and $[\,\cdot\,]_{H_1}$
denotes the classes in $H(\Gr_\Sigma)/H_1$.
Since the restriction of $\sigma_\eg$ to its image is an automorphism and
$\ker\sigma_0=\ker\sigma^2_0$ we have
\be
H(\Gr_{\Sigma_\eg})
\cong H(\Gr_\Sigma)/\erz{\{H_1,R^\eg w_i,i=1,2\}}
\ee
$\Sigma_\eg$ denoting the graph having $\sigma_\eg$ as connectivity
matrix.
Any of the \sst s $\srho_\eg$ leads to a homeomorphism $Q_\eg$ between
$\Om_\Rh$ and the path space ${\cal P}_{\Sigma_\eg}$ and to a
surjective homomorphism $[Q^*_\eg]:H(\Gr_{\Sigma_\eg})\rightarrow H(\Gr)$.
Let
$\pi_\eg:H(\Gr_\Sigma)\to H(\Gr_{\Sigma_\eg})$ be the natural projection.
Then $[Q_4^*]=[Q_\eg^*]\circ \pi_\eg$ for all $\eg$ and therefore
\be
\ker [Q_4^*] \supset\erz{\{H_1,R^\eg w_i,i=1,2;\,\eg=1,\cdots,9\}}\cong\Z^{32}.
\ee
This shows independently that
$H(\Gr)\subset\Z^{8}$ but not the opposite inclusion.
A computation of $\mbox{im}\,\KK^{(4,4)}$ for e.g.\ $\rho_0$
would have been more complicated due to the lack of symmetry.

Connes associates to the Penrose tilings yet another graph, the folded
$A_4$ Coxeter graph, making use
of a coding of a tiling by $0,1$ sequences obeying the condition that
no consecutive $1$'s can appear \cite{Cone}. But this coding, which was found
by Robinson \cite{GrSh}, does not distinguish between a tiling and
its image under an element of the orientational symmetry. In fact,  the
coding yields a homeomorphism between the hull modulo the orientational
symmetry and the path space of the folded $A_4$ graph, with the effect that
the groupoid arising is that given by cofinality. In other words,
the tilings are considered
as equivalence classes under translations {\em and} rotations and reflections.
One then obtains an $AF$-algebra as groupoid-\CA\
whose $K_1$-group is trivial and whose $K_0$-group coincides with the
integer group of coinvariants of the groupoid
and may be identified with the group $I$ in (\ref{03083})
as ordered group. Although the range of the tracial state on it
coincides
with (\ref{03082}) up to the order unit (the factor $\frac{1}{20}$
does not appear) it is a priori not clear that it predicts the right
gap labels since the $AF$-algebra does not contain the
discrete Laplacian.

\subsection*{Concluding Questions}
\addcontentsline{toc}{section}{\bf Concluding Questions}


We have computed the integer group of coinvariants and its order but
even in two dimensions and under the assumption that
$K_0(\A_\tc)=H(\Gr)\oplus \Z$ it is not clear what the order structure
on the $K_0$-group is.\ms

We did not mention groupoid cohomology but at least for
tilings which reduce to decorations the integer group of coinvariants
is a cohomology group of the groupoid $\Gr$ with coefficients in $\Z$.
The result of \cite{FoHu} on the connection between $K$-theory and
group cohomology
is easily seen to generalize to this situation since the stability of
$K$-theory under taking stably isomorphic algebras
is mirrored by
the stability of
(continuous) groupoid cohomology under taking
(continuously) similar groupoids. It is therefore tempting to
believe that $K$-theory of algebras associated to tilings is always related to
the groupoid cohomology with coefficients in $\Z$ of the associated
groupoid in a way like in  \cite{FoHu}.

 \newpage
\be \label{19011}
N^{0} =
\left(\begin{array}{c}
3\;0\;0\;0\;0\;0\;0\;0\;0\;0\;1\;0\;0\;0\;1\;0\;2\;0\;0\;0\;0\;0\;0\;0\;0\;0\;0
\;1\;0\;0\;1\;0\;0\;0\;0\;0\;
   0\;0\;0\;0\\ 0\;1\;0\;0\;0\;0\;0\;0\;0\;1\;0\;0\;0\;0\;0\;2\;0\;0\;0\;0\;0
\;0\;0\;0\;0\;1\;0
\;0\;0\;0\;
   0\;0\;0\;0\;0\;0\;0\;0\;1\;0\\
0\;0\;0\;0\;0\;0\;0\;0\;0\;0\;0\;0\;0\;0\;0\;0
\;0\;0\;0\;0\;0\;0\;0\;0\;
   0\;0\;0\;0\;0\;0\;0\;0\;0\;0\;0\;0\;0\;0\;0\;0\\
   0\;0\;0\;0\;0\;0\;0\;0\;0\;0\;0\;0\;0\;0\;0\;0\;0\;0\;0\;0\;0\;0\;0\;0\;0
\;0\;0
\;0\;0\;0\;0\;0\;0\;0\;0\;0\;
   0\;0\;0\;0\\ 1\;0\;0\;0\;1\;0\;1\;0\;0\;0\;1\;0\;0\;0\;0\;0\;1\;0\;0
\;0\;1\;0\;0\;0\;0
\;0\;0\;0\;0\;0\;
   0\;0\;0\;0\;0\;0\;0\;0\;0\;0\\
0\;0\;0\;0\;0\;3\;0\;0\;0\;0\;0\;2\;0\;0\;0\;1
\;0\;0\;0\;1\;0\;0\;1\;0\;
   0\;0\;0\;0\;0\;0\;0\;0\;0\;0\;0\;1\;0\;0\;0\;0\\
   0\;0\;0\;0\;1\;0\;1\;0\;0\;0\;2\;0\;0\;0\;0\;0\;0\;0\;0\;0\;1\;0\;0\;0\;0
\;0\;0\;0\;0\;0\;0\;0\;0\;1\;0\;0\;
   0\;0\;0\;0\\ 0\;0\;0\;0\;0\;0\;0\;0\;0\;0\;0\;0\;0\;0\;0\;0\;0\;0\;0\;0
\;0\;0\;0\;0\;0\;0
\;0\;0\;0\;0\;
   0\;0\;0\;0\;0\;0\;0\;0\;0\;0\\
0\;0\;0\;0\;0\;0\;0\;0\;0\;0\;0\;0\;0\;0\;0\;0
\;0\;0\;0\;0\;0\;0\;0\;0\;
   0\;0\;0\;0\;0\;0\;0\;0\;0\;0\;0\;0\;0\;0\;0\;0\\
   0\;1\;0\;0\;0\;1\;0\;0\;0\;1\;0\;1\;0\;0\;0\;1\;0\;0\;0\;0\;0\;0\;0\;0\;0
\;1\;0\;0\;0\;0\;0\;0\;0\;0\;0\;0\;
   0\;0\;0\;0\\ 1\;0\;0\;0\;2\;0\;1\;0\;0\;0\;3\;0\;0\;0\;0\;0\;0\;0\;0\;0
\;1\;0\;0\;0\;0\;0
\;0\;0\;0\;0\;
   0\;0\;0\;1\;0\;0\;0\;0\;0\;0\\
0\;0\;0\;0\;0\;1\;0\;0\;0\;1\;0\;1\;0\;0\;0\;1
\;0\;0\;0\;1\;0\;0\;0\;0\;
   0\;0\;0\;0\;0\;0\;0\;0\;0\;0\;0\;1\;0\;0\;0\;0\\
   0\;0\;0\;0\;0\;0\;0\;0\;0\;0\;0\;0\;0\;0\;0\;0\;0\;0\;0\;0\;0\;0\;0\;0\;0\;0
\;0\;0\;0\;0\;0\;0\;0\;0\;0\;0\;
   0\;0\;0\;0\\ 0\;0\;0\;0\;0\;0\;0\;0\;0\;0\;0\;0\;0\;0\;0\;0\;0\;0\;0\;0
\;0\;0\;0\;0\;0
\;0\;0\;0\;0\;0\;
   0\;0\;0\;0\;0\;0\;0\;0\;0\;0\\
2\;0\;0\;0\;0\;0\;0\;0\;0\;0\;0\;0\;0\;0\;1\;0
\;1\;0\;0\;0\;0\;0\;0\;0\;
   0\;0\;0\;1\;0\;0\;1\;0\;0\;0\;0\;0\;0\;0\;0\;0\\
   0\;1\;0\;0\;0\;1\;0\;0\;0\;2\;0\;0\;0\;0\;0\;3\;0\;0\;0\;0\;0\;0\;0\;0\;0
\;1\;0\;0\;0\;0\;0\;0\;0\;0\;0\;0\;
   0\;0\;1\;0\\ 1\;0\;0\;0\;1\;0\;0\;0\;0\;0\;1\;0\;0\;0\;1\;0\;1\;0\;0\;0
\;0\;0\;0\;0\;0
\;0\;0\;0\;0\;0\;
   1\;0\;0\;0\;0\;0\;0\;0\;0\;0\\
0\;0\;0\;0\;0\;0\;0\;0\;0\;0\;0\;0\;0\;0\;0\;0
\;0\;0\;0\;0\;0\;0\;0\;0\;
   0\;0\;0\;0\;0\;0\;0\;0\;0\;0\;0\;0\;0\;0\;0\;0\\
   0\;0\;0\;0\;0\;0\;0\;0\;0\;0\;0\;0\;0\;0\;0\;0\;0\;0\;0\;0\;0\;0\;0\;0\;0
\;0\;0\;0\;0\;0\;0\;0\;0\;0\;0\;0\;
   0\;0\;0\;0\\ 0\;0\;0\;0\;0\;2\;0\;0\;0\;0\;0\;1\;0\;0\;0\;0\;0\;0\;0\;1
\;0\;0\;1\;0\;0\;0
\;0\;0\;0\;0\;
   0\;0\;0\;0\;0\;1\;0\;0\;0\;0\\
1\;0\;0\;0\;0\;0\;0\;0\;0\;0\;0\;0\;0\;0\;1\;0
\;1\;0\;0\;0\;0\;0\;0\;0\;
   0\;0\;0\;0\;0\;0\;1\;0\;0\;0\;0\;0\;0\;0\;0\;0\\
   0\;0\;0\;0\;0\;0\;0\;0\;0\;0\;0\;0\;0\;0\;0\;0\;0\;0\;0\;0\;0\;0\;0\;0\;0\;0
\;0\;0\;0\;0\;0\;0\;0\;0\;0\;0\;
   0\;0\;0\;0\\ 0\;1\;0\;0\;0\;0\;0\;0\;0\;1\;0\;0\;0\;0\;0\;2\;0\;0\;0\;0
\;0\;0\;0\;0\;0\;1
\;0\;0\;0\;0\;
   0\;0\;0\;0\;0\;0\;0\;0\;1\;0\\
1\;0\;0\;0\;1\;0\;0\;0\;0\;0\;1\;0\;0\;0\;1\;0
\;1\;0\;0\;0\;0\;0\;0\;0\;
   0\;0\;0\;0\;0\;0\;1\;0\;0\;0\;0\;0\;0\;0\;0\;0\\
   0\;0\;0\;0\;0\;0\;0\;0\;0\;0\;0\;0\;0\;0\;0\;0\;0\;0\;0\;0\;0\;0\;0\;0\;0\;0
\;0\;0\;0\;0\;0\;0\;0\;0\;0\;0\;
   0\;0\;0\;0\\ 0\;0\;0\;0\;0\;1\;0\;0\;0\;0\;0\;1\;0\;0\;0\;0\;0\;0\;0\;1\;0
\;0\;0\;0\;0\;0\;0
\;0\;0\;0\;
   0\;0\;0\;0\;0\;1\;0\;0\;0\;0\\
0\;0\;0\;0\;0\;0\;0\;0\;0\;0\;0\;0\;0\;0\;0\;0
\;0\;0\;0\;0\;0\;0\;0\;0\;
   0\;0\;0\;0\;0\;0\;0\;0\;0\;0\;0\;0\;0\;0\;0\;0\\
   0\;0\;0\;0\;1\;0\;1\;0\;0\;0\;2\;0\;0\;0\;0\;0\;0\;0\;0\;0\;1\;0\;0\;0\;0
\;0\;0
\;0\;0\;0\;0\;0\;0\;1\;0\;0\;
   0\;0\;0\;0\\ 0\;0\;0\;0\;0\;1\;0\;0\;0\;1\;0\;1\;0\;0\;0\;1\;0\;0\;0\;1
\;0\;0\;0\;0\;0\;0
\;0\;0\;0\;0\;
   0\;0\;0\;0\;0\;1\;0\;0\;0\;0\\
0\;0\;0\;0\;0\;0\;0\;0\;0\;0\;0\;0\;0\;0\;0\;0
\;0\;0\;0\;0\;0\;0\;0\;0\;
   0\;0\;0\;0\;0\;0\;0\;0\;0\;0\;0\;0\;0\;0\;0\;0\\
   0\;0\;0\;0\;1\;0\;1\;0\;0\;0\;1\;0\;0\;0\;0\;0\;0\;0\;0\;0\;1\;0\;0\;0\;0\;0
\;0\;0\;0\;0\;0\;0\;0\;0\;0\;0\;
   0\;0\;0\;0\\ 0\;0\;0\;0\;0\;0\;0\;0\;0\;0\;0\;0\;0\;0\;0\;0\;0\;0\;0\;0\;0
\;0\;0\;0\;0\;0\;0
\;0\;0\;0\;
   0\;0\;0\;0\;0\;0\;0\;0\;0\;0\\
0\;1\;0\;0\;0\;1\;0\;0\;0\;1\;0\;1\;0\;0\;0\;1
\;0\;0\;0\;0\;0\;0\;0\;0\;
   0\;1\;0\;0\;0\;0\;0\;0\;0\;0\;0\;0\;0\;0\;0\;0\\
   2\;0\;0\;0\;0\;0\;0\;0\;0\;0\;0\;0\;0\;0\;1\;0\;1\;0\;0\;0\;0\;0\;0\;0\;0
\;0\;0
\;1\;0\;0\;1\;0\;0\;0\;0\;0\;
   0\;0\;0\;0\\ 0\;0\;0\;0\;0\;0\;0\;0\;0\;0\;0\;0\;0\;0\;0\;0\;0\;0\;0\;0
\;0\;0\;0\;0\;0\;0
\;0\;0\;0\;0\;
   0\;0\;0\;0\;0\;0\;0\;0\;0\;0\\
0\;1\;0\;0\;0\;0\;0\;0\;0\;1\;0\;0\;0\;0\;0\;1
\;0\;0\;0\;0\;0\;0\;0\;0\;
   0\;1\;0\;0\;0\;0\;0\;0\;0\;0\;0\;0\;0\;0\;0\;0\\
   0\;0\;0\;0\;0\;0\;0\;0\;0\;0\;0\;0\;0\;0\;0\;0\;0\;0\;0\;0\;0\;0\;0\;0\;0\;0
\;0\;0\;0\;0\;0\;0\;0\;0\;0\;0\;
   0\;0\;0\;0\\ 1\;0\;0\;0\;1\;0\;1\;0\;0\;0\;1\;0\;0\;0\;0\;0\;1\;0\;0\;0
\;1\;0\;0\;0\;0\;0
\;0\;0\;0\;0\;
   0\;0\;0\;0\;0\;0\;0\;0\;0\;0\\
0\;0\;0\;0\;0\;2\;0\;0\;0\;0\;0\;1\;0\;0\;0\;0
\;0\;0\;0\;1\;0\;0\;1\;0\;
   0\;0\;0\;0\;0\;0\;0\;0\;0\;0\;0\;1\;0\;0\;0\;0\\
   0\;0\;0\;0\;0\;0\;0\;0\;0\;0\;0\;0\;0\;0\;0\;0\;0\;0\;0\;0\;0\;0\;0\;0\;0\;0
\;0\;0\;0\;0\;0\;0\;0\;0\;0\;0\;
   0\;0\;0\;0
\end{array}\right)
\ee
\newpage
\epsffile[0 0 465 615]{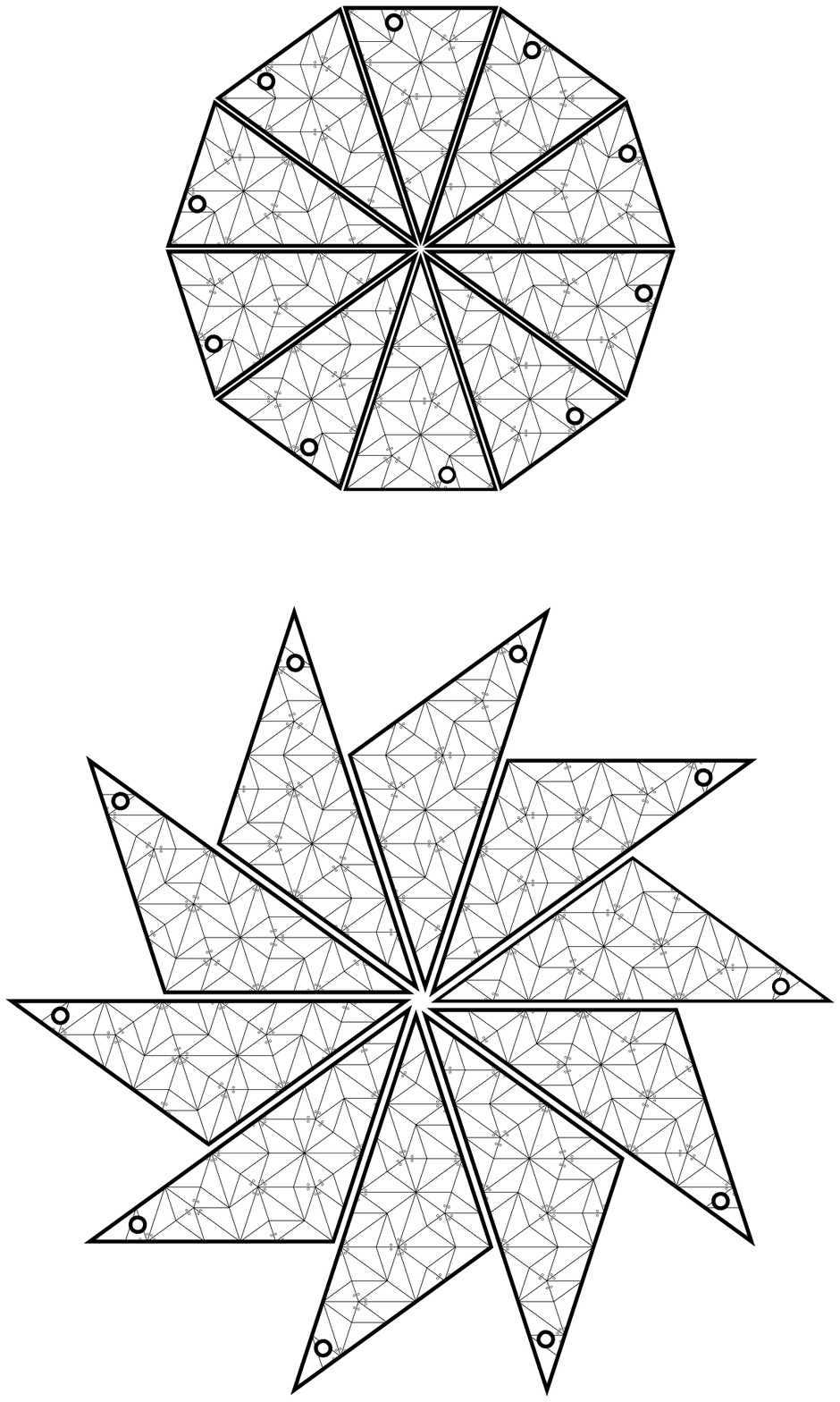}
Figure 8: $4$-fold substitutes of the triangles of Figure~7.0 and
7.1 in all directions.
\newpage
\epsffile[0 0 465 645]{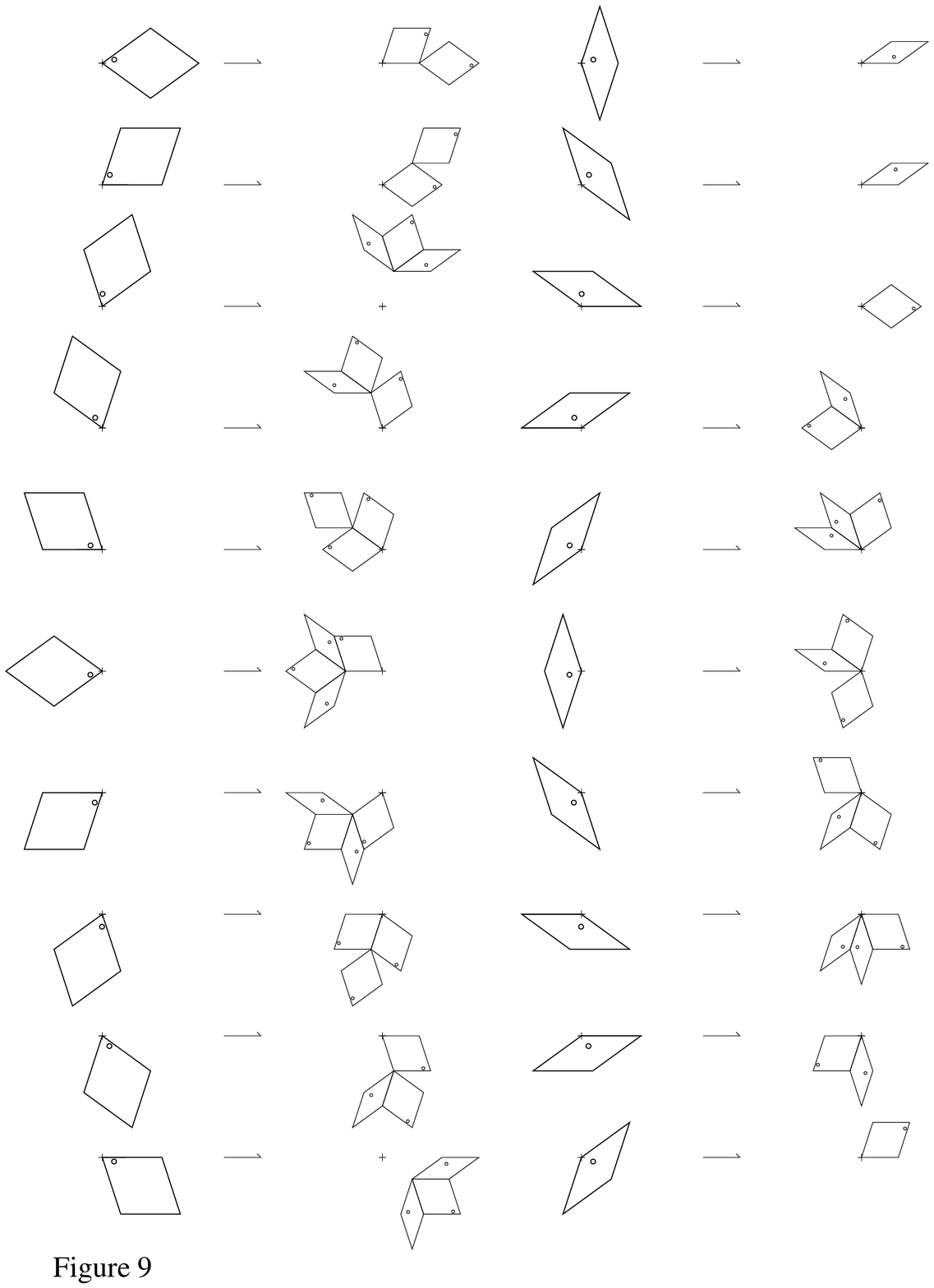}
\newpage
\addcontentsline{toc}{section}{\bf References}


\newpage
\tableofcontents

\end{document}